\documentclass[fleqn,usenatbib]{mnras}

\usepackage{newtxtext,newtxmath}
\usepackage[T1]{fontenc}
\usepackage{ulem}
\DeclareRobustCommand{\VAN}[3]{#2}
\let\VANthebibliography\thebibliography
\def\thebibliography{\DeclareRobustCommand{\VAN}[3]{##3}\VANthebibliography}

\usepackage{graphicx}	% Including figure files
\graphicspath{{images/}}
\usepackage{amsmath}	% Advanced maths commands
\title[NGC\,6752 WD\,CS with \textit{HST}]{
The \textit{HST} large programme on NGC 6752 – IV. The White Dwarf Sequence} 
\author[L.\ R.\ Bedin et al.]{
L.\, R.\, Bedin$^1$\thanks{E-mail: luigi.bedin@inaf.it}, 
M.\, Salaris$^2$, 
J.\, Anderson$^3$, 
M.\, Scalco$^{4,1,5,6}$,
D.\, Nardiello$^{1}$, 
E.\ Vesperini$^{6}$,
H.\ Richer$^{7}$,
\newauthor
A.\ Burgasser$^8$, 
M.\ Griggio$^{4,1}$, 
R.\, Gerasimov$^8$,
D.\ Apai$^{9,10}$,
A.\, Bellini$^3$,    
M.\, Libralato$^3$,
P.\ Bergeron$^{11}$, 
\newauthor
R.\ M.\ Rich$^{12}$, and 
A.\ Grazian$^{1}$  
\\
$^{1}$Istituto Nazionale di Astrofisica, Osservatorio Astronomico di Padova, Vicolo dell'Osservatorio 5, Padova I-35122, Italy\\
$^{2}$Astrophysics Research Institute, Liverpool John Moores University,146 Brownlow Hill, Liverpool L3 5RF, UK\\
$^{3}$Space Telescope Science Institute, 3800 San Martin Drive, Baltimore, MD 21218, USA\\ 
$^{4}$Dipartimento di Fisica, Universit\`a di Ferrara, Via Giuseppe Saragat 1, Ferrara I-44122, Italy\\
$^{5}$ESO, Karl-Sc hwarzsc hild Str asse 2, D-80 Garc hing, Germany\\
$^{6}$Department of Astronomy, Indiana University , Swain Hall West, 727 E 3rd Street, Bloomington, IN 47405, USA\\
$^{7}$Department of Physics and Astronomy, University of British Columbia, Vancouver BC V6T 1Z1, Canada\\
$^{8}$Center for Astrophysics and Space Science, University of California San Diego, La Jolla, CA 92093, USA\\
$^{9}$Department of Astronomy and Steward Observatory, The University of Arizona, 933 N. Cherry Avenue, Tucson, AZ 85721, USA\\
$^{10}$Lunar and Planetary Laboratory, The University of Arizona, 1640 E. University Blvd., Tucson, AZ 85721, USA\\
$^{11}$D\'epartement de Physique, Universit\'e de Montr\'eal, C.P.\,6128, Succ.\,Centre-Ville, Montr\'eal, QC\,H3C\,3J7, Canada\\
$^{12}$Department of Physics and Astronomy, UCLA, 430 Portola Plaza, Box 951547, Los Angeles, CA 90095-1547, USA\\
}
\date{Accepted 2022 November 4. Received 2022 November 3; in original form 2022 September 12}
\pubyear{2022}
\begin{document}
\label{firstpage}
\pagerange{\pageref{firstpage}--\pageref{lastpage}}
\maketitle

\begin{abstract}
We present our final study of the white dwarf cooling sequence
(WD\,CS) in the globular cluster NGC\,6752.  The investigation is the main goal of a
dedicated \textit{Hubble Space Telescope} 
large Program,
for which all the observations are now collected.  
The WD\,CS luminosity function (LF)
is
confirmed to peak at $m_{\rm F606W} \simeq 29.3 \pm 0.1$, consistent
within uncertainties with what has been previously reported, and is
now complete down to $m_{\rm F606W} \simeq 29.7$.  
We have performed robust and conclusive comparisons with
model predictions that show how the theoretical LF for hydrogen
envelope WD models closely follow the shape of the empirical LF. The
magnitude of the peak of the observed LF is matched with ages between
12.7 and 13.5 Gyr, consistent with the cluster age derived from the
main sequence turn off and subgiant branch.  We also find that the
impact of multiple populations within the cluster on the WD LF for
$m_{\rm F606W}$ below 27.3 is negligible, and that the presence of a
small fraction of helium envelope objects is consistent with the data.
Our analysis reveals a possible hint of an underestimate of the
cooling timescales of models in the magnitude 
range 28.1 $<$ $m_{\rm F606W}$ $<$ 28.9. 
Finally, we find that hydrogen envelope models calculated with a new
tabulation of electron conduction opacities in the transition between
moderate and strong degeneracy provide WD ages that are too small in
comparison to the Main Sequence turnoff age.
\end{abstract}

\begin{keywords}
white dwarfs -- globular clusters: individual: NGC\,6752. 
\end{keywords}

%%%%%%%%%%%%%%%%%%%%%%%%%%%%%%%%%%%%%%%%%%%%%%%%%%

%%%%%%%%%%%%%%%%% BODY OF PAPER %%%%%%%%%%%%%%%%%%

%%%%%%%%%%%%%%
\section{Introduction}
%

%%%
\begin{figure*}
    \centering
    \includegraphics[width=\textwidth]{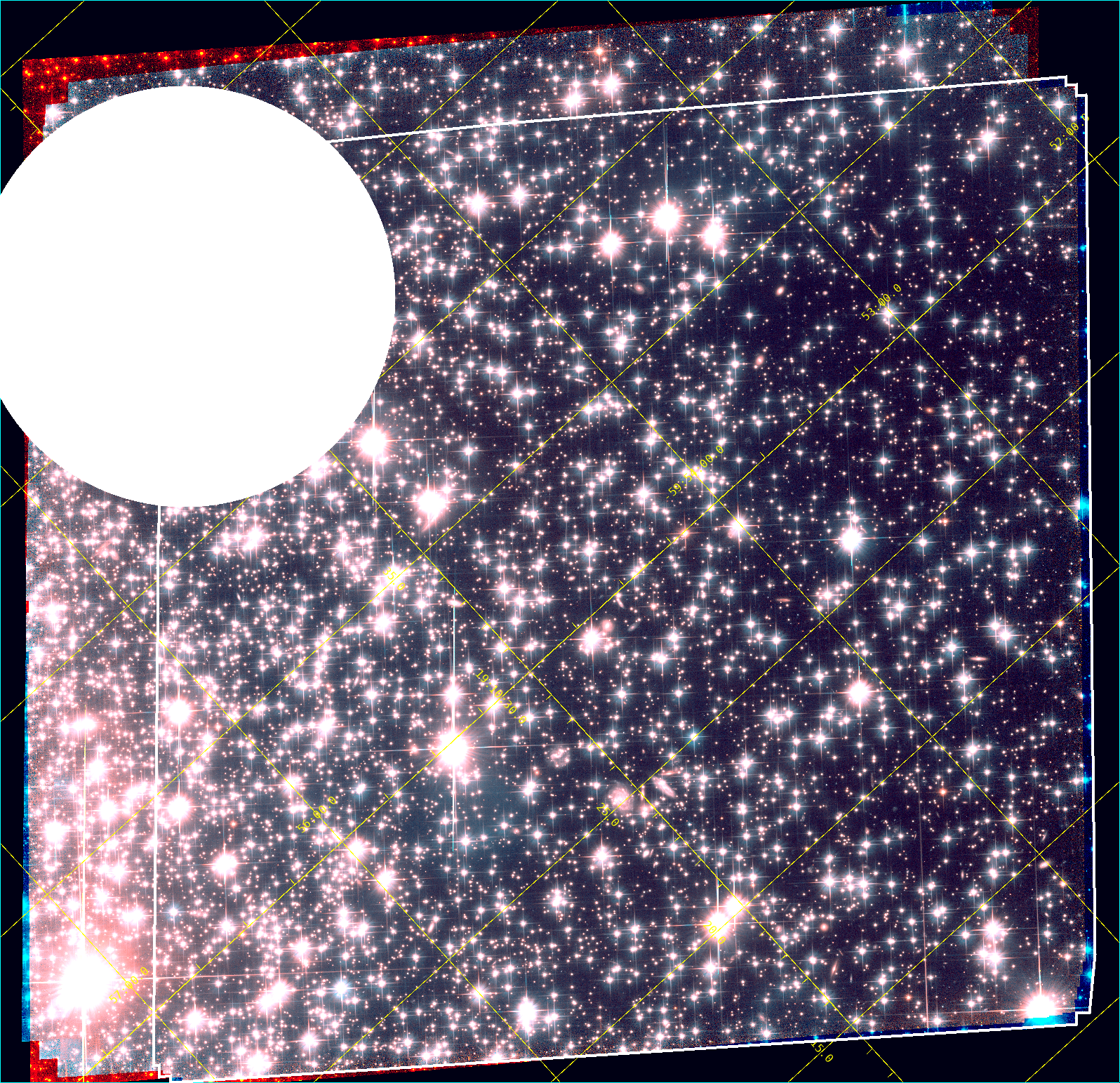}
    \caption{
%%%%    
Tri-chromatic view of about $250^{\prime\prime} \times
250^{\prime\prime}$ for the entire ACS/WFC data-set from \textit{HST}
programs GO-15096 \& GO-15491.  The white lines indicate the region
observed at first epoch.  The circular white region is rejected from
this work, as it used to mask out the \textit{Bedin\,I} dwarf
spheroidal galaxy in background (Paper\,I).  The yellow lines and
labels indicate the \textit{Gaia}\,DR3 reference frame.
%%%%    
    }
    \label{fov}
\end{figure*}

Located at a distance of about 4~kpc \citep{bv}, NGC\,6752 is one of
the closest Milky Way globular clusters (GCs) to the Sun.  With its
proximity and low interstellar reddening, NGC\,6752 was one of the
first GCs to reveal the presence of white dwarfs (WDs) \citep{richer}
before the advent of \textit{Hubble Space Telescope (HST)}. 
A small sample of bright WDs in NGC\,6752 observed with \textit{HST}
was employed by \citet{rb} to determine the cluster distance with a
technique analogous to the main sequence fitting.  More recently,
\citet{bsa} using deeper \textit{HST} observations, obtained a
first, preliminary, WD luminosity function (LF) that reached the end of
the cooling sequence of hydrogen envelope WDs.  Thus, NGC\,6752
belongs to an exclusive club of GCs for which the end of the WD
cooling sequence has been observed, including NGC\,6397
\citep{r6397, h6397}, M\,4 \citep{M4} and 47\,Tuc \citep{hTuc}.

One peculiarity of NGC\,6752 is the atypical horizontal branch (HB)
morphology for its metallicity ([Fe/H]$\sim-$1.6,
\citealt{gratton}). There are no RR~Lyrae stars in its HB, which has a
pronounced extension to the blue, related to the presence of multiple
populations with different initial helium abundances \citep[e.g.,][and
  references therein]{bl18, csp14}.  This blue morphology affects the
bright part of the WD cooling sequence (CS), as shown very recently by
\citet{chen22} based on near-UV imaging observations with \textit{HST}
Wide Field Camera\,3. Using synthetic HB modelling by \citet{csp14}
and theoretical results on WD progenitor evolution by \citet{althaus},
\citet{chen22} have shown that the bright part of the CS is populated
by two types of WDs. A \lq{slow\rq} --\,in terms of evolutionary
speed\,-- component supported by envelope hydrogen burning, and a
\lq{canonical}\rq \, component which is cooling without any
substantial contribution from nuclear burning (more on this later).

This work is the fourth of a series of articles aimed at the
scientific exploitation of an \textit{HST} multi-cycle large program
focused on the GC NGC\,6752.  The previous three publications were
based on data from the first epoch of observations.
\citet[][Paper\,I]{2019MNRAS.484L..54B} presented the discovery of a
dwarf spheroidal galaxy that needs to be masked out in studies of the
faint population of NGC\,6752.
\citet[][Paper\,II]{2019MNRAS.484.4046M} characterized the multiple
stellar populations of NGC\,6752; Finally,
\citet[][Paper\,III]{bsa} reached, for the first time in NGC\,6752,
the end of the hydrogen envelope WD CS.

In this study we have employed data for all epochs, which
significantly improves the definition of the cluster WD CS and its LF,
and allows us to perform a more robust and reliable investigation of
the agreement between WD and main sequence turn off ages, the effect
of the cluster multiple populations on the faint end of the CS, and
the impact of helium envelope WDs on the LF. We have also been able to
test two different treatments of the electron conduction opacities in
the regime between moderate and strong degeneracy, which are crucial
for the modelling of WD envelopes and have a major impact on the
cooling times of the models \citep[see][]{opaconf}.

The article is organized as follows:
Section~\ref{obs} presents the observations, while 
section\,\ref{datareduction} gives details for the data reduction.
Section~\ref{artificial} provides a brief description of the artificial star tests.
Section~\ref{cmd} describes the selection criteria used to obtain the cluster colour magnitude diagram (CMD).   
Section~\ref{sec:pms} explains the decontamination of the cluster sample using proper motions.
Section~\ref{wdcslf} presents the derived empirical WD LF.
Section~\ref{model} compares the empirical LF with theory.
Section~\ref{conclusions} summarizes our results.

%%%%%%%%%%%%%%
\section{Observations}
\label{obs}
This study is based on images collected with the \textit{Wide Field
  Channel} (WFC) of the \textit{Advanced Camera for Surveys} (ACS) on
board \textit{HST} under the multi-cycle large program: \textit{"The
  end of the White Dwarf Cooling Sequence of NGC\,6752"}, programs:
GO-15096 and GO-15491 (PI: Bedin).  Paper\,III presented the first
epoch of data obtained as part of program GO-15096, collected between
September 7 and 18, 2018.  These data consist of deep exposures of
$\sim$1270\,s each, 19 taken with the F814W filter and 56 taken with
the F606W filter.  Short exposures of $\sim$45\,s each were also
collected at the beginning of each orbit, 10 with the F814W filter and
27 with the F606W filter. Five of the planned 40 orbits failed (some
only partially) because of poor guide-star acquisition and were
repeated between August 1 and 15, 2019, resulting in an additional
5$\times$45\,s and 10$\times$1270\,s exposures with the F814W filter.

The second half of the data were collected between September 2 and 11,
2021 as a part of GO-15491.  Due to changes in the
\textit{HST}-phase-II policies, deep exposures were on average shorter
by $\sim$55\,s to allow for more ease in the scheduling.
We obtained long exposures of $\sim$1215\,s each, 56 images with the
F814W filter and 20 images with the F606W filter; and short exposures
of $\sim$45\,s, each 12 with the F814W filter and 28 images with the
F606W filter.  Two orbits were lost due to poor guiding, and these
observations were repeated on February 14, 2022; with two short
(45\,s) and four deep (1209\,s) exposures with the F814W filter.
All images were collected between $\sim$2018.68 and $\sim$2022.12,
resulting in four epochs over a time-baseline of $\sim$3.5\,yrs.  In a
forthcoming publication, we use these multi-epoch observations to
conduct an astrometric analysis of stars bright enough to be detected
in individual images ($V \lesssim 28$) to determine the absolute
motions, parallax, and internal velocity dispersion of the cluster.
Here, we analyze all of the images simultaneously to detect the
faintest WD members of the cluster.
%%%

%%%%%%%%%%%%%%
\section{Data-reduction}
%\label{artificial}
\label{datareduction}
The data reduction was essentially identical to that presented in
Paper\,III but with about twice the number of images. We refer the
interested reader to Paper\,III and previous publications for details
on the procedures, and provide here a brief description.

We downloaded from the MAST archive\footnote{\texttt{mast.stsci.edu}}
the \texttt{flc} images, which were pre-processed with \textit{Space
  Telescope Science Institute (STScI)}'s pipeline. The \texttt{flc}
images are corrected for dark current, bias, flat fielding, and
charge-transfer efficiency (CTE) losses (following the
\citealt{2010PASP..122.1035A} recipes for pixel-based correction) with
the latest reference files, but with no re-sampling of the pixels.

We first conducted a \textit{``first-pass''} analysis to derive
optimized point-spread functions (PSFs) for all images, and to
establish a common distortion-free reference frame.  Fluxes and
positions for relatively bright (down to $\sim$3.5\,magnitudes below
saturation), unsaturated stars were extracted from each \texttt{flc}
image using software developed by J.\ Anderson, described in
\citet{2006acs..rept....1A}.
Each image was analyzed separately to create a tailored PSF in order
to account for the particular breathing state of \textit{HST}'s
telescope tube, which affects both spatial and temporal variations
relative to the library PSFs provided by \citet{2006hstc.conf...11A}.
This tailoring of PSFs was done with prescriptions introduced in
\citet{2017MNRAS.470..948A} for WFC3/UVIS, and later extended to
ACS/WFC by \citet{Bellini2018}. 
Next, both positions and
fluxes are corrected for the geometric distortion of the detector by
\citet{2006hstc.conf...11A},\footnote{
Publicly available at \texttt{https://www.stsci.edu/$\sim$jayander/\-HST1PASS/LIB/GDCs/STDGCs/}
}  
which affects pixel areas and hence fluxes.  These geometric
corrections were used to produce a common, distortion-free reference
frame --based on cluster members-- to which all individual images are
linked.  Note that during the $\sim$3.5\,years between the first and
last epoch, cluster members are expected to have internal motions on
the order of $\sim$1\,mas; field objects will have much larger motions
(more about motions in Sect.\,\ref{sec:pms}).
% Bellini et al, 2018 ACS/WFC ISR 2018-08
% \bibitem[Bellini et al.(2018)]{2018acs..rept....8B} Bellini, A., Anderson, J., \& Grogin, N.~A.\ 2018, Instrument Science Report ACS 2018-8

This \textit{``first-pass''} analysis yields a distortion-free
reference frame, with positions accurate to milli-arcsecond (mas)
levels, and magnitudes zero-pointed to milli-magnitude (mmag)
precision levels.  With these calibrations in hand, we performed a
\textit{``second-pass''} analysis in which all of the pixels from all
of the images are analyzed simultaneously to search for the faintest
sources in the field, in particular those not detectable in individual
images.  This analysis was done using the most recent version of the
code (\texttt{KS2}) developed by J.\, Anderson (first presented in
\citealt{2008AJ....135.2055A}), and applied in several other GC
analyses (cf.\ \citealt{2021MNRAS.505.3549S} and references therein).
The \texttt{KS2} code goes through multiple iterations of finding,
modelling, and subtracting point sources from the image, starting from
the brightest sources and moving progressively to fainter sources in
the subtraction residuals.  The code solves for positions and fluxes,
as well as other important diagnostic parameters such as the local sky
noise (\texttt{rmsSKY}) which documents how noisy the investigated
patch of sky is, and the \texttt{RADXS} parameter (introduced in
\citealt{M4}) which documents how well the source flux distribution
resembles that of the PSF.  The \texttt{RADXS} parameter is the most
efficient diagnostic to eliminate faint unresolved galaxies, poorly
measured stars perturbed by non-modellable neighbors, cosmic ray (CR)
hits, PSF substructure, diffraction spikes, and other artifacts.
Stars that are saturated in the deep exposures have valid first-pass
measurements from the short exposures and are linked to the
deep-exposure-based master frame via common unsaturated stars.

Photometry was calibrated to the ACS/WFC Vega-mag system with the
procedure given in \citet{b05} using encircled energy and zero points
available from STScI.\footnote{
  www.stsci.edu/hst/acs/analysis/zeropoints} For the derived
magnitudes in this photometric system, we adopt the
symbols $m_{\rm  F606W}$ and $m_{\rm F814W}$.

%%%%%
The \texttt{KS2} code also produces stacked images for visual
examination, which we release as part of the electronic material
associated to this work.  We combined these stacks for the two filters
to produce a color view of the studied region, shown in
Fig.\,\ref{fov}.  [A tri-chromatic view was obtained using F606W \&
  F814W for the blue \& red channels, while using a
  wavelength-weighted mean of F606W and F814W for the green channel].
In the figure we masked out a region centered on the dwarf spheroidal
galaxy \textit{Bedin\,I} (coordinates from Paper\,I) with a radius of
1000\,pixels ($\sim50^{\prime\prime}$). This region is completely
excluded from our analysis, as it proved impossible to achieve a
useful discrimination in the CMD between the location of the dwarf
galaxy's stars and the WDs of NGC\,6752.

\begin{figure}
    \centering
    \includegraphics[width=\columnwidth]{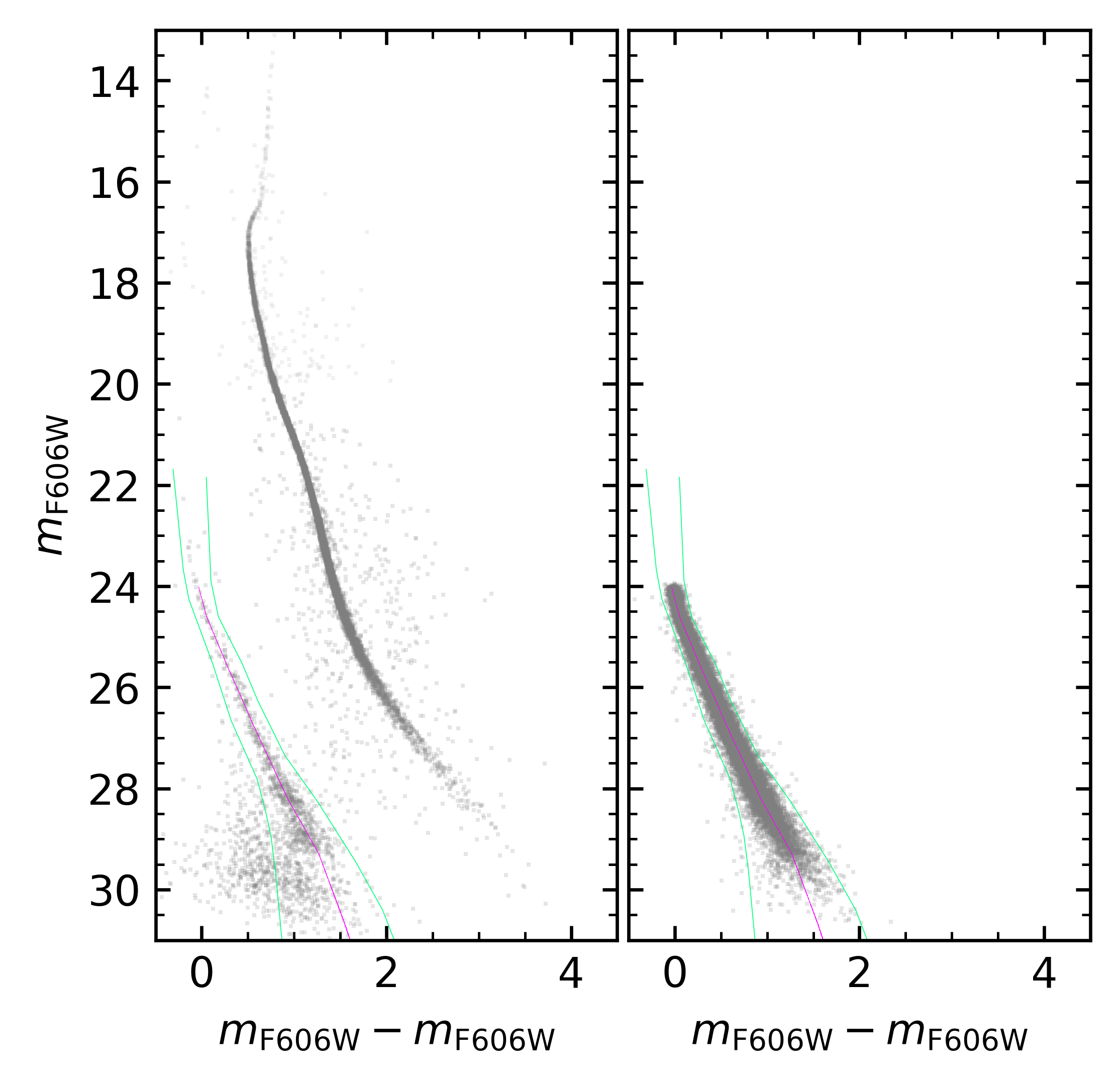}
    %%%%
    \caption{
    %%%%
%
    \textit{(Left:) } Preliminary CMD for the sources within the studied field. 
    Only sources with $\texttt{RADXS}\,<0.1$ in both F606W and F814W are displayed. 
    The line in magenta, obtained as described in Sect.\,\ref{artificial}, 
    is the fiducial line for the WD\,CS of NGC\,6752. 
    The green lines define the region of the CMD within which we will count the WDs. 
    \textit{(Right:)} With the same scale, 
    we show the CMD for artificial stars. The same magenta and green lines of the left panel
    are also displayed here. 
    As it can be seen, the lines in green define a region that is generous enough 
    to include both the bulk of the observed real WDs in the left panel \textit{and} 
    the artificial stars added along the WD fiducial line recovered with  
    large photometric errors.
    Note that the exact location of the magenta line does not affect the number of 
    WDs counted within the region defined by the two green lines.
    %%%%
    }
    %%%%

%
\label{fig:1stCMD}
\end{figure}

The absolute astrometric registration to the International Celestial
Reference System (ICRS) was achieved using sources in common with
\textit{Gaia} Early Data Release 3 (EDR3;
\citealt{2021A&A...649A...1G}). Tabulated proper motions in
\textit{Gaia}\,EDR3 were transformed to the average Julian Day of
images collected during the first epoch, following the procedures in
\citet{2018MNRAS.481.5339B}.\\~
%%%%%%
%
In the left panel of Fig.\,\ref{fig:1stCMD} we show a preliminary CMD for 
sources having $|$\texttt{RADXS}$|$\,$<0.1$ in both F606W and F814W. 
In the next section, we will employ this CMD to define the 
fiducial line of the WD\,CS of NGC\,6752. This fiducial will be used to generate
artificial stars along it, which in turn will enable us to define 
the region within which to count the WDs of NGC\,6752, and 
to carefully select well-measured stars (see Sect.\,\ref{cmd}). 
%%%
%%%

%%%%%%%%%%%%%%
\section{Artificial Stars}
\label{artificial}
When studying faint sources, artificial star tests (hereafter, ASTs)
have several key roles, specifically:
\textit{(i)} ASTs are used to track and correct for systematic errors
between input magnitude and recovered magnitudes (see \citealt{M4});
\textit{(ii)} ASTs are also employed to estimate the random errors and
therefore define which sources have a position on the CMD consistent
with being WDs;
\textit{(iii)} ASTs are used to check and define the selections on the
distribution of diagnostic parameters, such as \texttt{RADXS} or
\texttt{rmsSKY};
and finally --and of fundamental importance-- \textit{(iv)} ASTs are
used to assess the completeness of the sample.
In creating the ASTs, the first step is to choose where to add them 
in the CMD, and where to add them spatially across the field of view. 
Since we are studying the WD\,CS of NGC\,6752, on the CMD we define a fiducial line, 
drawn by hand, along the bulk of the observed WDs down to 
where they seem to stop, and extrapolate the fiducial to even fainter magnitudes 
in order to assess completeness.
This fiducial line is shown in the left panel of Fig.\,\ref{fig:1stCMD} 
(in magenta, as in other figures of this article).  
With \texttt{KS2}, we added artificial stars along this fiducial line
with a flat distribution in $m_{\rm F606W}$ between magnitudes 24 and 32, 
and with a homogeneous spatial distribution across the field of view.  
Therefore, this fiducial line was defined on the observed WD\,CS of NGC\,6752. 
We did not use theoretical models, nor we made assumptions about the WD\,CS location. 
We follow the prescriptions in \citet[section 2.3]{M4},
and correct our magnitudes (both real and artificial) for input-output
systematic errors, which are negligible at $m_{\rm F606W}\sim$24, but
become as large as $\sim$0.2\,mag at the faintest magnitudes. In the
following, our magnitudes --for both artificial and real sources-- are
corrected for these effects.

Unfortunately, ASTs cannot track down all the possible sources of
photometric systematic errors, the most important being related to CTE
effects.
Indeed, in the case of real sources, their photo-released charge is
bitten out by the electron traps encountered in the detector that are
filled during the read-out process, all the way down to the amplifier
(up to $\sim$2000 pixels).
ASTs instead are just added values to the analyzed images, and their
artificial charge was never subjected to all the electron-traps which
would lay along the read out process.

The recent ISR by Anderson et al.\, (in preparation) use an observing
strategy to self-calibrate these CTE-related photometric errors, which
requires multiple observations of the same field to be collected at
$\sim$90 degrees from each other.
As our observations of NGC\,6752 where not collected this way, a
correction of this kind here is not possible.
However, results in that work suggest that in the
case of our data set, any \textit{possible} systematic photometric
error related to the CTE losses, could amount to $\sim$0.1\,mag which
therefore in the following we set as the limit of our photometric
accuracy.
We also note that F606W exposures have on average a 45$\%$ higher
sky-background level than those in F814W ($\sim$136 vs.\ $\sim$94
photo-electrons) making photometry F606W less vulnerable to CTE
losses, as many more electron traps are filled.
Furthermore, the number of images in F606W is more than twice the
number available in F814W (precisely 112 vs 53), making
$m_{\rm F606W}$-magnitudes more robust measurements than those in F814W.
For this reason, in the following, we will chose to analyze both the
CMD and the LF in the $m_{\rm F606W}$-magnitudes.

To establish whether an inserted star was recovered or not,
we assumed that if an artificial star is not recovered within 0.753
($\sim -2.5\log{2}$) magnitude (in both filters), and within 1\,pixel
from the inserted position (in both $x$ and $y$ detector coordinates),
then the inserted artificial star was not recovered.
The right panel of Fig.\,\ref{fig:1stCMD} shows artificial sources as inserted
(in magenta) and how they were recovered (dots). 
The panels of Fig.\,\ref{fig:1stCMD}, together, are used to define the 
region within which we will count the WDs of NGC\,6752. 
Specifically, this region is defined by two lines (in green) drawn by hand, 
which are a compromise between including the observed WDs of NGC\,6752  
with large photometric scatter, and excluding the vast majority of field objects. 
We allowed for photometric errors very large around the magenta line, 
so that the bulk of the observed WDs (which might not follow exactly the magenta line), 
will still be included. 
%

%%%%%%%%%%%%%%

\begin{figure*}
    \centering
    \includegraphics[width=\textwidth]{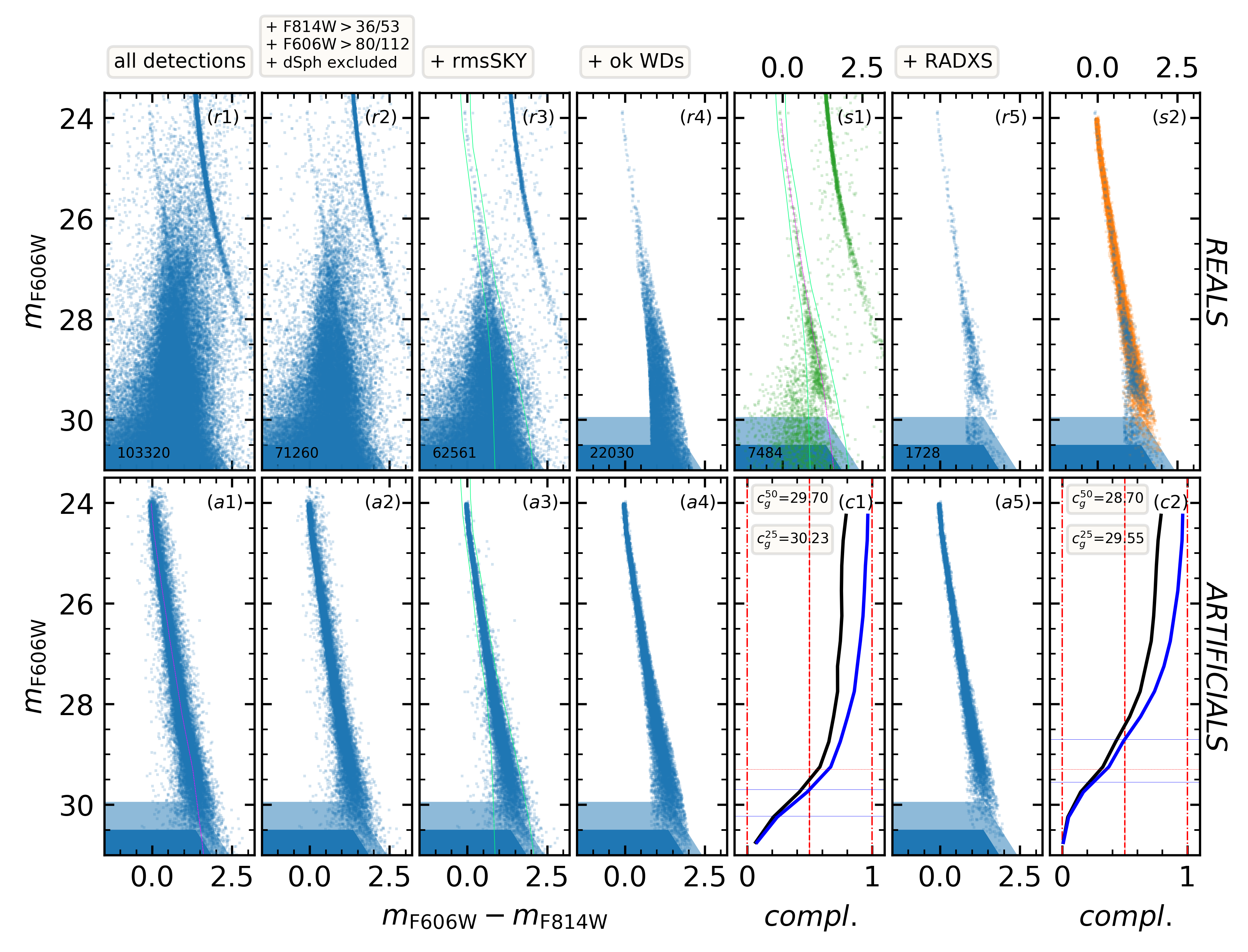}
    \caption{
%%%%    
Panels show the progression of the cumulative selections adopted to
obtain a sample of well-measured WDs along NGC\,6752's WD CS.  From
left to right panels from $(1)$ to $(5)$ for real stars [top $(r_i)$
  panels] and artificial stars [bottom $(a_i)$ panels]. A thin line in
magenta in panel $(a1)$ show where artificial stars were added.
Panels $(c)$ show the obtained completeness, not including $(c1)$ or
including $(c2)$ the selection on RADXS, which is by far the most
effective parameter to select well measured point-sources.  For
clarity in panel $(s1)$ we "show" the effect of RADXS selection on
stars also outside the WD region defined by the two green thin lines
(see text). Finally, in panel $(s2)$ we show the direct comparison of
real vs.\ artificial stars, clearly not showing a sharp drop in their
number below $m_{\rm F606W}\simeq 29.4$.
%%%%    
    }   
    \label{sel}
\end{figure*}

%%%%%%%%%%%%%%
\section{Colour-Magnitude Diagram and Selections}
\label{cmd}
Following the approach of Paper\,III, in Fig.\,\ref{sel} we show the
impact of our progressive selection criteria on artificial stars, and
then apply the criteria identically to real sources.  Each panel is
labelled in the top-right corner with an $(a)$ for ASTs, or with an
$(r)$ for real sources.  The goal of these selections is to find the
best compromise between keeping the largest sample of well-measured WD
members of NGC\,6752 and rejecting the detections that are most likely
spurious, or poorly measured stars, or field objects in the foreground
and background of the cluster -- be they either stars or galaxies.

The \texttt{KS2} code sets a detection threshold for potential sources
to be any positive local maximum in an image that is above 1-$\sigma$
of the local sky noise.  Naturally, this choice results in the
inclusion of a large number of non-significant detections, but it has
the value of showing where the floor-level of the white noise is
located.  As in Paper\,III, we measured the background noise
(1-$\sigma$) in the two filters for regions with average low
background and transformed these values into magnitudes, associating
to these peaks the value of the PSFs' central-pixel value (normalized
to unity), and zero-pointing to the Vega-mag system.  We show in
panels $(a)$ and $(r)$ with shading the corresponding 5-$\sigma$
(light-blue) and 3-$\sigma$ (dark blue) regions.  In the following, we
will not consider as significantly detected any source below the
5-$\sigma$ limits. Therefore, sources above these shaded regions could
still be poorly measured objects or even artifacts, but nevertheless
they are solid detections.

Panel \textit{(a1)} show all the artificial sources as inserted (in magenta)
and as they were recovered (blue dots).

In panel \textit{(a2)} we restricted our sources to those that fell in
the part of the field that was observed in at least $\sim$70$\%$ of
the F814W and F606W images. Given our large dither observing strategy,
this leads to a significant reduction of the field of view that is
used for this investigation.

Panel \textit{(a3)} further restricts us to a region of the field
where the \texttt{rmsSKY} was consistent with the noise within empty
patches of sky, i.e., to regions that were not dominated by the bright
halo of luminous stars. This selection has only a marginal effect in
rejecting sources in noisy regions of the FoV, but it has a great
significance in establishing which regions are suitable
(\textit{"good"}) for the detection of faint objects (thanks to the
lower sky noise level).  The computed completeness in \textit{"good"}
regions, is indicated by $c_g$, following \citet{2008ApJ...678.1279B}.

In Panel \textit{(a4)} we use the region enclosed within  
the two green lines in panel \textit{(a3)} to reject all the ASTs recovered 
outside this region.

In the next panel, \textit{(c1)}, instead of a CMD, we show the
magnitude vs.\ completeness curve (i.e., the: number-of- recovered /
number-of-inserted sources) with a black line, and the completeness
limited to "good" regions ($c_g$) as a blue line for the surviving
artificial stars.  This panel shows that inserted sources that passed
these selections are 50\%-complete down to $m_{\rm F606W}\simeq 29.7$,
and 25\%-complete at $m_{\rm F606W}\simeq 30.2$. However, we can
recognize that while 50\%-completeness is well above the sky noise at
5-$\sigma$, the 25\%-completeness is well below our 5-$\sigma$ minimum
threshold for significant sources.  In the following, completeness
corrections are assumed to be reliable only for magnitudes brighter
than the 5-$\sigma$ magnitude level expected for sky noise, even
though we show some points below this.

In the next panel \textit{(a5)}, we show the result after selection
with our most effective diagnostic to reject non-stellar objects,
i.e., the \texttt{RADXS} parameter.  This parameter is able to reject
most of the PSF artifacts, diffraction spikes, extended sources, and
field objects that have moved significantly more than the cluster
member stars (about 1\,mas) during the $\sim$3.5 years between first
and last epochs, causing a blur of the shape of stars that do not move
as cluster members\footnote{Since cluster members are used to compute
  the transformations, they do not move within errors, especially
  toward faint magnitudes, where random errors dominate
  uncertainties.}.

The bottom-right panel \textit{(c2)} shows the completeness curves
after the final selection on \texttt{RADXS} is also applied.  Note
that 25\%-level of the $c_g$-completeness remains well above the
5-$\sigma$ sky-level down to magnitudes $m_{\rm F606W}\simeq 29.55$,
i.e., below the location of the WD\,CS LF's peaks as observed in
Paper\,III.\\

In the top panels of Fig.\,\ref{sel}, we have applied the very same
selections defined for ASTs to the observed real sources.  In panel
\textit{(r5)}, we show the final sample for NGC\,6752's WD candidates.
We also \textit{show} the selection of two interesting real-star CMDs (labelled
with \textit{(s)}).  

Panel \textit{(s1)} shows the CMD for stars that
passed all the selection criteria, but includes sources falling
outside the region between the 
two thin green lines defined in panel \textit{(a3)}.
This CMD includes non-WD candidates. Sources in this CMD are coloured in green, and will be used in
Sect.\,\ref{sec:pms} to derive a model of the field contamination
within the WD region in the CMD.
Note also that the fiducial line defined in Sect.\,\ref{artificial} (and shown 
in magenta) well represents the mean observed CMD location for WD\,CS of NGC\,6752. 

Panel \textit{(s2)} shows a direct comparison of the CMDs for sources
surviving all the selection criteria: ASTs (as selected in panel
\textit{(a5)}) are shown in orange, while real sources (as selected in
panel \textit{(r5)}) are shown in blue.  From this CMD, it is evident
that the observed real WD\,CS does not extend to magnitudes as faint
as those of the recovered ASTs --- a clear indication that we have
reached and passed the peak of the WD\,CS LF of NGC\,6752.

%%%%%

\begin{figure*}
    \centering
    \includegraphics[width=\textwidth]{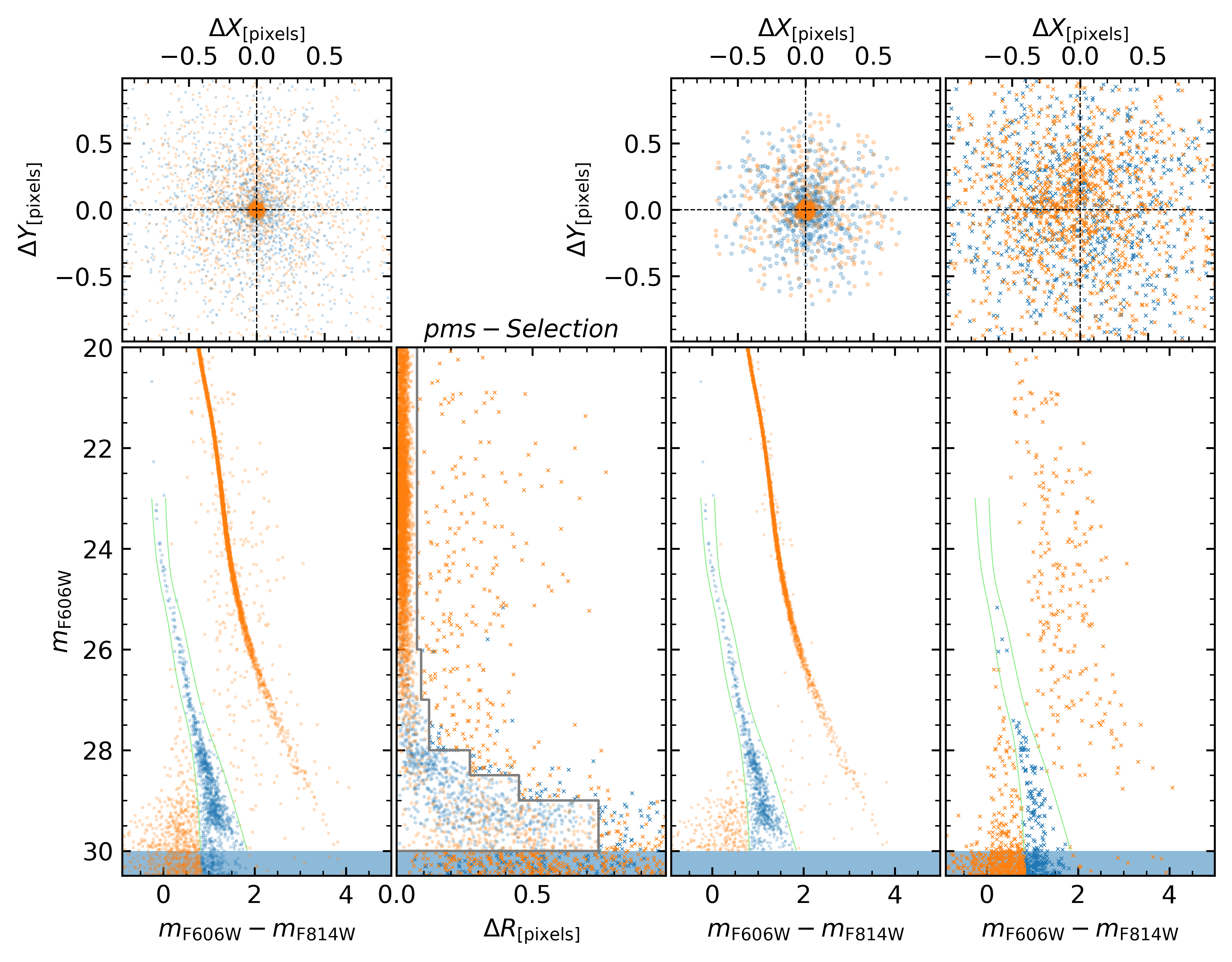}
    \caption{
    \textit{(Bottom-left:) } CMD of the sources in panel (s1) of Fig.\,\ref{sel},    
    here sources within the WD-region (between the two green lines)
    are color-coded in blue and all the other sources in
    orange. Shading below $m_{\rm F606W}=30$ marks the limit of
    significant detection for the sources of interest.
    \textit{(Bottom-mid-left:)} 1-D displacements, $\Delta\,R$, as
    function of $m_{\rm F606W}$. Bright stars (down to $m_{\rm F606W}
    \sim 28$) show a $\Delta\,R$ distribution with a tight dispersion
    ($<0.1$\,pixels) plus a tail with a much broader dispersion peaked
    around 0.3\,pixels.
    We arbitrary define two regions, indicated by the gray line, one
    that enclose the bulk of the $\Delta\,R$ at different magnitude
    bins (dots), and the other with objects of larger $\Delta\,R$
    (small crosses).
    The next two bottom-panels on the right show the CMDs for the
    stars within and beyond the gray line.
    Clearly neither of the two CMDs are made only of members or only of field objects (see text). 
    \textit{(Top:) } 2-D vector point diagrams of the $\sim$3\,yr-displacements for the 
    samples shown in the corresponding panels below. 
    }
    \label{fig:pms}
\end{figure*}

%%%%%%%%%%%%%%
\section{Proper-motion decontamination}
\label{sec:pms}

We now check whether the proper-motions (PMs) derived for the faintest
stars, can be of any help to discriminate between field objects
falling within the WD region of the NGC\,6752's CMD, and true cluster
members.  To this end, we combined the bulk of the first half of the
data collected in $\sim$2018.7 and obtain averaged positions for the
sources, which are then compared with their averaged positions as
measured in the second half of the data collected in $\sim$2021.7.  In
the following analysis, we will consider only sources shown in green
in panel \textit{(s1)} of Fig.\,\ref{sel} (i.e., those surviving all
the selection cuts described above, therefore considering sources
outside the region of the WDs in the CMD), but for which it was
possible to estimate a position at each of the two considered epochs.

Unfortunately, the absolute motion of NGC\,6752 (relative to the
field) is not very large:
$(\mu_{\alpha\cos{\delta}};\mu_\delta)=(-3.155;-4.010)\pm(0.008;0.009)$\,mas\,yr$^{-1}$
(\citealt{2022ApJ...934..150L}), resulting in a combined absolute
motion of just $5.102\pm0.012$\,mas\,yr$^{-1}$.  With a time-baseline
of just $\sim$3\,yrs we can expect a separation of about $\sim$15\,mas
between the cluster members and the faintest extra-galactic unresolved
field sources that sit in an absolute rest frame.  Given the ACS/WFC
pixel-size of 49.72\,mas, this amounts to a displacement of about
0.3\,pixel, which is an easily measurable quantity for the brightest
stars, but it is smaller than the measurements errors for those
extremely faint sources that become significant only when combining
several dozen individual images of the same epoch.

We provide a quantitative illustration of this situation in
Fig.\,\ref{fig:pms}.  We color-code in blue stars surviving the WD
selection defined by the two thin green lines, and in orange all other
stars.
In the top panels, we show vector-point diagrams of source
displacements over the $\sim$3\,years, in units of ACS/WFC pixels (of
49.72\,mas) for 2D-displacements ($\Delta X, \Delta Y$) between the
two epochs (2018.7 vs.\ 2021.7). In the first, third and fourth
bottom-panels we show CMDs.  The second bottom-panel from left, shows
the observed 1D-displacement ($\Delta R$ summing in quadrature the
displacements in $X$ and in the $Y$ axes) as a function of the
observed source $m_{\rm F606W}$-magnitude.  Among bright stars, it is
clear that there is a tight distribution in $\Delta R$ for cluster
members, which remain distributed well below 0.1\,pixel, and a broad
tail toward higher $\Delta R$, peaking between 0.2-0.4 pixels: these
are field objects.  However, at around magnitude $m_{\rm F606W} \simeq
28$, the random positional-measurement errors (which are summed in
quadrature for the two epochs) explode for fainter stars.  And at
around magnitude $m_{\rm F606W} \simeq 28.5$, it clearly becomes
impossible to disentangle members (with positional random errors
around 0.4\,pixels) from field objects (with a relative average
displacement of $\sim$0.3\,pixels).

With the observed bulk of stars at different magnitude intervals, we
define the PM selections for objects consistent with the PM errors at
the various magnitudes.  This arbitrary selection is illustrated by
the thick-grey line, Stars satisfying this selection (grey step-line)
are indicated with circles, while stars not passing this selection or
fainter than $m_{\rm F606W} \simeq 30$ are indicated with crosses.
The panels on the right of this panel, show the vector-point diagram
and the CMD for sources to the left or to the right of the gray-line
criterion, keeping the blue color-code for WD candidates.  The PM
selection, while useful to reject outliers and objects with large PMs,
is completely useless at separating WDs and field objects below
$m_{\rm F606W} \simeq 28.5$, simply because the separation between
field and members is much smaller than measurement errors below
$m_{\rm F606W} \simeq 28.5$.  In the next section, we introduce a
work-around to properly define the WD\,CS LF of NGC\,6752 down to its
faintest magnitudes.

%%%

\begin{figure*}
    \centering
    \includegraphics[width=\textwidth]{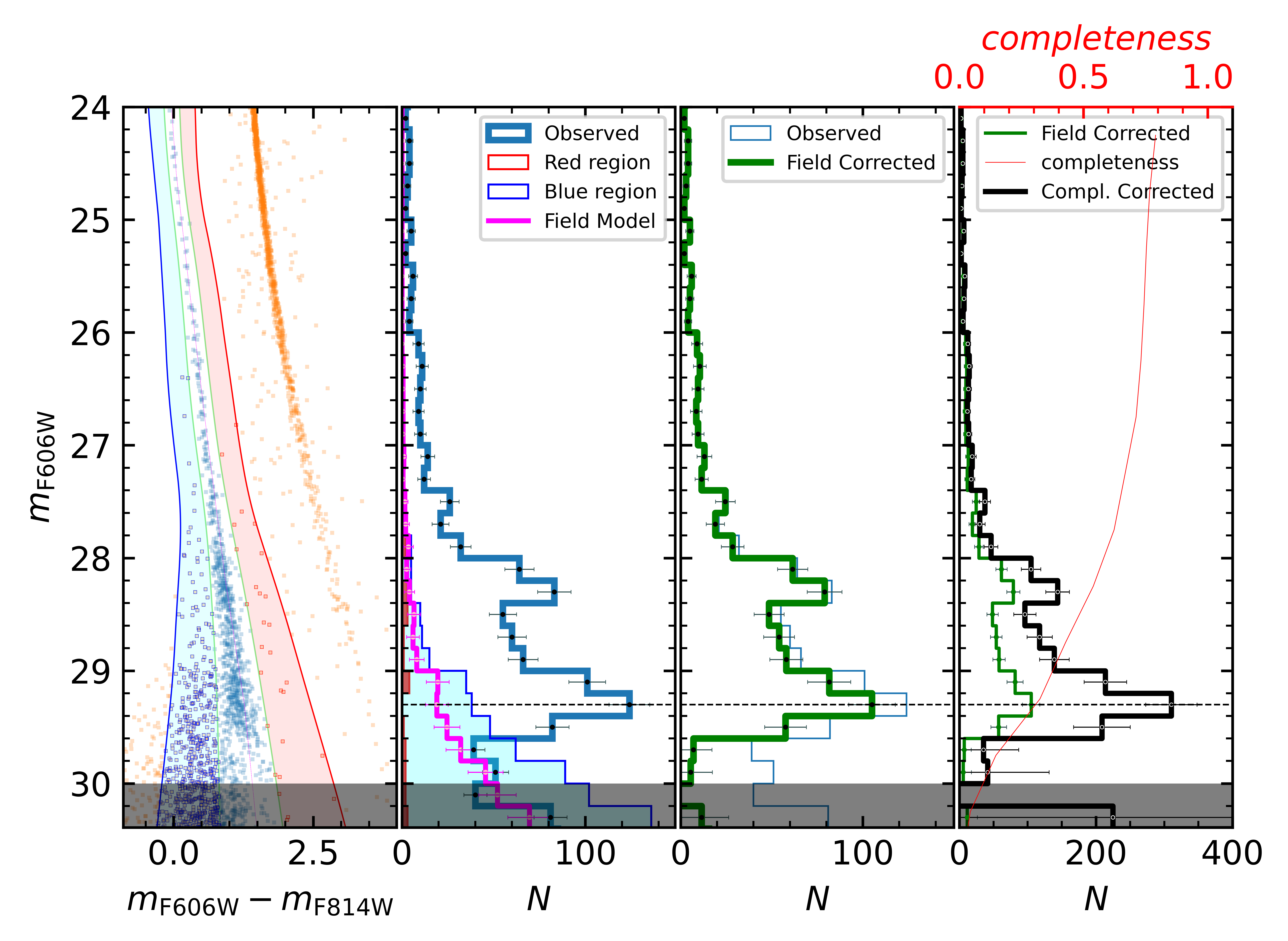}
    \caption{\textit{(Left:) } The CMD where we define three regions
      along the WD\,CS.  The shaded region in azure indicates sources
      passing the selections on the blue side of the WDs, the region
      shaded in red, the ones redder.  Next panels are all WD CS LFs,
      where the histograms show the number of sources per magnitude
      interval, for observed stars within the WD region and for stars
      in the two shaded regions (see the legend).  Our model for field
      distribution is the histogram in magenta.  This is a simple
      average of the blue and red LF.  This simple model is then
      subtracted from the observed LF, and shown in the third panel
      from left to right.  Note how the faintest bins of the corrected
      WD\,CS\,LF are consistent with zero within the noise.  Finally,
      the observed field-corrected WD\,CS's LF was corrected for
      completeness, and shown in black.  Grey shaded regions indicate
      levels where findings, and completeness becomes unreliables (see
      text).  Errors where linearly propagated and then corrected for
      compleness.
    }
    \label{fig:wdcslf}
\end{figure*}

%%%%%%%%%%%%%%
\section{The corrected WD CS LF}
\label{wdcslf}

As demonstrated in various \textit{HST} studies, including those that
made use of the \textit{Hubble Ultra Deep Field} (e.g.,
%\citealt{M4,2010ApJ...708L..32B}),
\citealt{2008ApJ...678.1279B,M4,2010ApJ...708L..32B}), the vast
majority of contaminants in the CMD aligned with the cluster's WD CS
are blue galaxies that are relatively easy to reject with shape
parameters such as RADXS. However, at fainter magnitudes, unresolved
blue galaxies become increasingly indistinguishable from stars.
Although these faint unresolved blue point-sources fall in a region
just outside the WD\,CS (e.g., \citealt{M4} and discussion, or Fig.\,3
of Paper\,III), a number of blue stars in the Galactic field
contaminate the LF. Unfortunately, this contamination appears far from
negligible, and for a reliable study of the WD\,CS \,LF of NGC\,6752
some correction of residual contaminants should be performed.

In this section we develop a simple model to correct for field
contamination.  The process is illustrated in Fig.\,\ref{fig:wdcslf}
and described in the following.  In the first panel of this figure,
from left-to-right we show the CMD of sources defined in panel
\textit{(s1)} of Fig.\,\ref{sel}, and use the two lines of
Fig.\,\ref{sel} to define on the CMD what we will refer to as the
\textit{"WD-region"}.  We then defined two other regions with the
identical color-width at each magnitude of the CMD WD-region, but one
at bluer colors (to the left of the WD-region), and one at redder
colors (to the right of the WD-region).
In the following, we refer to these as \textit{"Blue-"} and
\textit{"Red-regions"}.  Next, we counted the sources within each of
these three regions, and produced in the next panel to the right the
luminosity functions (LFs) of each.  We visualize the LF for those WD
candidate objects observed within the WD-region, and the Blue- and
Red-regions, with error bars that reflect statistical Poisson errors.

In the simplest model for the contaminant distribution on the CMD's
WD-region, we assume that: \textit{(i)} there are no WDs of NGC\,6752
within the Blue- or Red-region; and \textit{(ii)} the number of
contaminants within the CMD WD-region is the average of the number of
objects observed in the Blue- and in the Red-regions, at the various
magnitudes.  We show this model in Fig.\,\ref{fig:wdcslf}, with
corresponding errors estimated by linear propagation of Poisson noise.

In the third panel of Fig.\,\ref{fig:wdcslf}, we compare the observed
LF to the resulting WD LF corrected for the field-contamination model.
The field-corrected LF is simply the observed LF minus the field
model, with errors propagated linearly.  Interestingly, in the
faintest bins, $m_{\rm F606W}$ = 29.6 and 30, the field correction
brings the observed WD\,CS\,LF to zero within the uncertainties.  As
this is what one would expected when the LF drops out --in spite of
the na\"ive assumption for this distribution of contaminants-- this is
a rather reassuring feature of the goodness and validity of the model.
Finally, in the last panel to the right, we show the
completeness-corrected and field-corrected WD CS LF for NGC\,6752 
The errors on this LF were also corrected for completeness with a 
simple approximation, a linear propagation of the errors.
As supplementary online material  we also release all data 
(completeness, observed LF, errors, CMD, photometric 
errors), to enable other groups to independently perform their own analyses. 
The WD\,LF in the right-panel of Fig.\,\ref{fig:wdcslf} is the one
used in the next Section for the theoretical analyses. 

\begin{figure*}
    \centering
    \includegraphics[width=\textwidth]{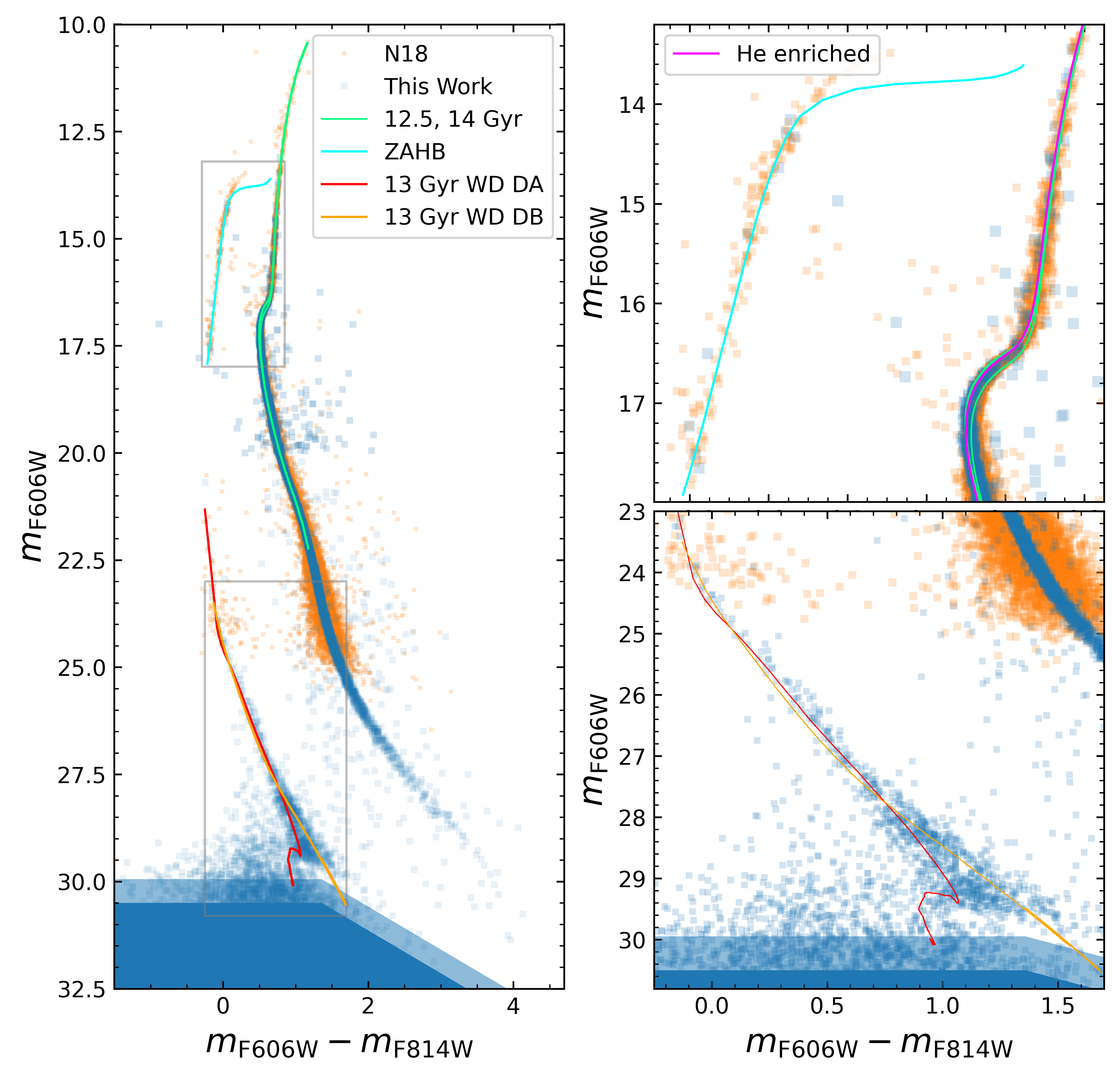}
    \caption{\textit{(Left:)} Cluster CMD compared to theoretical
      isochrones. In orange we display the catalogue from the central
      region of NGC\,6752 by \citet[N18]{2018MNRAS.481.3382N}, in blue
      the deep photometry for the external field studied in this work.
      The two sets of data agree; the richer central field of N18 is
      based on much shallower observations of a much more crowded
      region, so they become incomplete at $m_{\rm F606W} \sim 25$,
      but show well populated the bright and fast evolutionary phases,
      in particular: the RGB, the HB, and the bright part of the
      WD\,CS. The displayed ZAHB and theoretical isochrones are
      labelled. The 13~Gyr hydrogen envelope WD isochrone is denoted
      as DA, and the helium envelope counterpart is denoted as DB.
    \textit{(Right:)} zoom-in of the gray-rectangular regions
    indicated in the CMD of the left panel.  The \textit{Top} panel
    shows the MS+HB+RGB region with also the He-enriched isochrone (in
    magenta).  The \textit{Bottom} panel focus on the faint WD\,CS.
    }
    \label{fig:isofit}
\end{figure*}

%%%%%%%%%%%%%%
\section{Modelling the observed WD CS and LF}
\label{model}

As in Paper\,III, a preliminary step to model theoretically the
observed cooling sequence is to have estimates of the cluster
parameters, such as distance and reddening.
We employed here the same distance modulus $(m-M)_0$=13.10 and
reddening $E(B-V)$=0.05 as in Paper\,III; this distance is consistent
with the recent determination $(m-M)_0$=13.08$\pm$0.02 by \citet{bv},
while the reddening is consistent with $E(B-V)$=0.046$\pm$0.005
determined by \cite{gratton}.
We compared theoretical isochrones from the lower main sequence (MS)
to the tip of the red giant branch (RGB), and a zero age horizontal
branch sequence (ZAHB), to the optical CMD of the cluster central
regions from \citet[N18]{2018MNRAS.481.3382N}, merged with the deep
photometry for the external field studied in this work, as shown in
Fig.~\ref{fig:isofit}. These two photometries agree well, and the
richer central field populates the post main sequence phases and the
bright part of the WD\,CS.  We employed $\alpha$-enhanced
([$\alpha$/Fe]=0.4) isochrones and ZAHB models by \citet{bastialpha}
for [Fe/H]=$-$1.55 --consistent with [Fe/H]=$-1.55 \pm 0.01 \pm 0.06$
(random + systematic) determined by \citet{gratton}-- and an initial
helium mass fraction $Y$=0.248.  In this comparison with the observed
CMD (and also for the modelling of the WD sequence) we applied
extinction corrections to the $F606W$ and $F814W$ filters that depend
on the model effective temperature, calculated as in \citet{b05}.

Figure~\ref{fig:isofit} shows the overall good agreement between the
theoretical models and the observed sequences on the MS, RGB and ZAHB.
The plume of stars that departs from the ZAHB towards higher
luminosities is made of objects coming from the blue tail of the HB,
that are evolving towards the asymptotic giant branch (AGB).  This
comparison provides us with a MS turn-off (TO) age determination;
isochrones with ages equal to 12.5 and 14.0~Gyr match the TO region of
the CMD, and also bracket nicely the luminosity range of the subgiant
branch that is very sensitive to age.  We also display a 12.5~Gyr old
isochrone with initial $Y$=0.275 (keeping the metal content fixed) to
show the effect of the helium spread due to the presence of multiple
populations in the cluster.
Just as a reminder, the presence of multiple populations in a globular
cluster is revealed by (anti)correlated variations of the abundances
of C, N, O, Na (and sometimes Mg and Al), plus variations of the
initial helium mass fraction $Y$ \citep[see, e.g.,][]{r15}
A typical globular cluster like NGC\,6752 hosts a population of stars
with \lq{normal\rq} initial helium abundances and $\alpha$-enhanced
metal abundance patterns similar to those of field halo stars at the
same [Fe/H] --we denote it as first population-- together with a
population of helium-enriched stars displaying a range of N, Na and He
overabundance, coupled to C and O depletion compared to field stars at
the same [Fe/H] --we denote this as second population. When observing
in optical filters and as long as the sum of the CNO abundances is the
same in the first and second population --as generally from
spectroscopy-- stellar models calculated with the standard helium and
$\alpha$-enhanced composition of the first population are appropriate
to study observations of clusters' second populations \citep[see the
  review by][and references therein]{csreview20}.

In NGC\,6752, the average difference $\Delta Y$ among the cluster
multiple populations is estimated to be small, $\sim 0.02-0.03$
\citep{nardiello, m18, csp14}, and the $Y$=0.275 isochrone well
represents the cluster second population.  We can clearly see that the
effect of the enhanced He on the TO region of the isochrones is minor,
and does not appreciably affect the TO age estimates. Also, as
discussed by \citet{csp14}, the helium rich second population is
located towards the fainter blue end of the HB, a region where the
corresponding helium rich ZAHB would overlap with the $Y$=0.248 ZAHB
shown in Figure~\ref{fig:isofit}.

Moving now to the WD CS, Fig.~\ref{fig:isofit} also shows a comparison
between the observed CMD and two 13\,Gyr WD isochrones, employing the
same distance modulus and reddening as for the comparison to the MS,
RGB and ZAHB CMD.  The isochrones have been calculated from the
carbon-oxygen core WD cooling tracks by \citet{bastiwd}; specifically,
the models with hydrogen or helium envelopes and metal poor
progenitors \citep[see][for details]{bastiwd} computed with the
\citet{opacond} electron conduction opacities (see later for more
details on this). We employed the same initial-final mass relation
(IFMR) by \citet{ifmr} used to calculate the carbon/oxygen profiles of
the WD models, and the progenitor lifetimes from \citet{bastiwd}
evolutionary tracks with the appropriate metallicity ([Fe/H]=$-$1.55).
The helium envelope isochrone does not reach the faint end at this
age, because of limitations of the input physics of the models.  Due
to these limitations, the calculations of WD models above
$\sim$0.6$M_{\odot}$ could not reach cooling ages comparable with the
ages of GCs.
However, the fainter point along the helium envelope isochrone is
still much fainter than the magnitude of the cut off of the observed
LF, because of the faster cooling times compared to hydrogen envelope
models.

Figure~\ref{fig:isofit} displays an overall good agreement between
observations and isochrones from the bright to the faint end of the CS.
Below $m_{\rm F606W} \sim 28$, the hydrogen envelope isochrone becomes
increasingly offset towards bluer colours compared to observations, by
up to $\sim$0.1.  Given that hydrogen envelope WDs are present along
the cluster CS \citep[see, e.g.][]{rb, moehler, chen22} we believe
that systematic offsets of this order are compatible with the residual
uncertainties in the photometry of the faintest sources, that likely
affect mostly the F814W filter (see end of Sect.~\ref{artificial}).

Interestingly, the observed breadth of the faintest part of the WD\,CS
is larger than that inferred from ASTs, but bracketed by the two shown
isochrones.
There also seems to be some structure with a possible hint of a gap
around $m_{\rm F606W}\,\sim\,28.8$; however, only data with better
signal-to-noise would be able to investigate the exact shape of these
putative features.

%%%
\begin{figure}
    \centering
    \includegraphics[width=\columnwidth]{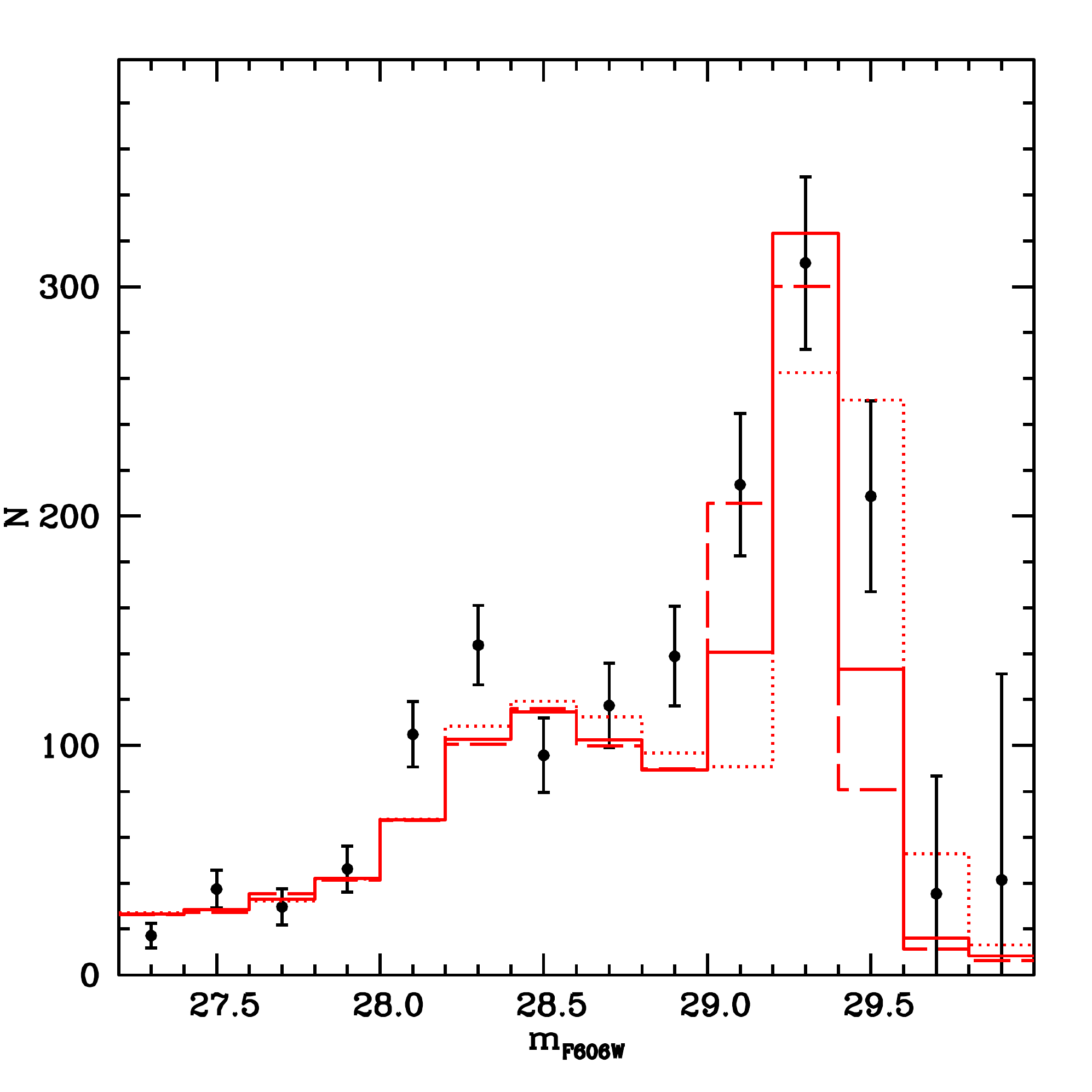}
    \caption{Observed WD LF compared to theoretical counterparts for
      ages equal to 12.7\,Gyr (dashed line), 13.0\,Gyr (solid line)
      and 13.5\,Gyr (dotted line) respectively. The theoretical LFs
      have been calculated using for the progenitors' mass function a
      power law with exponent $x$=$-$2.3, and are normalized by
      matching the total number of stars of the empirical LF between
      $m_{\rm F606W}$=27.3 and 27.9 (see text for details).}
    \label{fig:lfage}
\end{figure}
%%%

Our analysis of the WD LF focuses on isochrones from models with
hydrogen envelopes, as current evidence indicates that at least the
bright WDs in NGC\,6752 and other GCs are generally found with
hydrogen atmospheres \citep{moehler, rb, davis, chen22}. This
assumption is reinforced by the fact that, as we describe below, the
shape of the empirical LF agrees well with a population of WDs with
hydrogen envelopes.  We will briefly investigate the impact of the
possible presence of WDs with helium envelopes in Sect.~\ref{sec:He}

To study in more detail the consistency of the ages from the CS and
the MS TO, we need to model the WD LF. Before doing so, we need to
consider the effect of the HB morphology of this cluster on the WD
population.

As briefly discussed in the introduction, the HB of NGC\,6752 shows a
pronounced extension to the blue, with objects increasingly less
massive when moving towards bluer colours, because of thinner
hydrogen-rich envelopes around the helium core.
\citet{chen22} have discussed in detail how the bluer HB stars have
envelope masses too small to reach the asymptotic giant branch (AGB)
after the end of core helium burning, hence they do not experience the
thermal pulses and third dredge-up. These objects reach the WD stage
with a residual hydrogen envelope thick enough to sustain stable
thermonuclear burning \citep{althaus}. For redder HB stars with masses
above $\sim 0.56 M_{\odot}$ the post-HB evolution leads instead to the
AGB phase and the thermal pulses. The resulting WDs have thinner outer
hydrogen layers, of \lq{canonical\rq} mass thickness equal to $\approx
10^{-4} M_{\rm WD}$ where $M_{\rm WD}$ is the total WD mass, not able
to support efficient hydrogen burning \citep{althaus, chen21}.  As a
consequence, the bright part of the cluster CS is populated by two
different populations of WDs; a \lq{slower\rq} population supported by
envelope hydrogen burning, and a \lq{canonical\rq} population which is
cooling faster, without any substantial contribution from nuclear
burning.  The number ratio of these two populations is determined by
the mass distribution of the HB stars which, in turn, is controlled by
the distribution of initial helium abundances among the cluster's
multiple populations, as studied by \citet{csp14}.  Down to $m_{\rm
  F606W}\sim$25.1, the magnitude range studied by \citet{chen22}, this
number ratio \lq{slow\rq}/\lq{canonical\rq} is equal to $\sim 70/30$,
a value that is expected to progressively decrease at fainter
magnitudes because these WDs formed earlier in the life of the
cluster, when stars along the HB were more massive and increasingly
less blue.  At $m_{\rm F606W} \sim$ 27.3, corresponding to cooling
times of 3.-3.5~Gyr, we expect to have only \lq{canonical\rq} WDs
because their HB progenitors all had masses above 0.56$M_{\odot}$
\citep[see the discussion in][]{chen22}.

In our deep photometry, which covers the outer regions of the cluster,
the bright part of the LF is not well populated because of the sparser
density (necessary to study faint objects).  Therefore, we will need
to perform a comparison with model predictions considering
$m_{\rm F606W}\geq $27.3, where the number of objects is at least 20 per
bin.  Coincidentally, this is also the magnitude range where only
canonical DA WDs are expected to populate the CS.

We have calculated theoretical WD LFs starting from isochrones
computed as described before, using as reference a Salpeter-like power
law for the WD progenitor mass function with exponent $x$ set to
$-$2.3. The WD models we employ have negligible efficiency of Hydrogen
burning in their envelopes (the mass thickness of the hydrogen
envelope is equal to $10^{-4} M_{\rm WD}$). To properly include the
photometric errors we followed a Monte Carlo approach; we have first
randomly drawn progenitor masses according to the chosen mass
function, and determined the magnitude and colour of the corresponding
synthetic WD by interpolation along the isochrone, after applying the
distance modulus and extinction. The magnitude and colour have then
been perturbed by Gaussian random errors with 1\,$\sigma$ values
determined from the mean 1\,$\sigma$ errors at the magnitude of the
synthetic object, as obtained from the photometric analysis.  We
finally calculated the LF from the magnitude distribution of the
resulting sample of synthetic WDs.  For each LF calculation we have
drawn 100,000 synthetic stars, to minimize the statistical error in
the star counts, and normalized the theoretical LFs matching the total
number of objects in the empirical LF in the bins centred between
$m_{\rm F606W}$=27.3 and 27.9 (130 objects).

Figure~\ref{fig:lfage} shows theoretical LFs for the three ages that
match the position of the peak and subsequent drop of the star counts
at the faint end of the empirical LF. As well known, the magnitude of
this feature is a diagnostic for the total age (cooling age plus
progenitor age) of a cluster's WD population. The fainter the
magnitude, the older the WD population.  The range of ages inferred
from the theoretical LFs is between 12.7 and 13.5~Gyr, completely
consistent with the cluster age derived from the MS TO and subgiant
branch, as shown in Fig.~\ref{fig:isofit}.

The shape of the theoretical LFs follows closely the shape of the
empirical counterpart. In the magnitude range used for the
normalization, the theoretical LFs display a very mild increase of the
number counts with increasing magnitude, as observed.  There is a
sharper increase of the number counts toward fainter magnitudes, with
a roughly flat portion down to the bin centred at $m_{\rm
  F606W}$=28.9.

%%%
\begin{figure}
    \centering
    \includegraphics[width=\columnwidth]{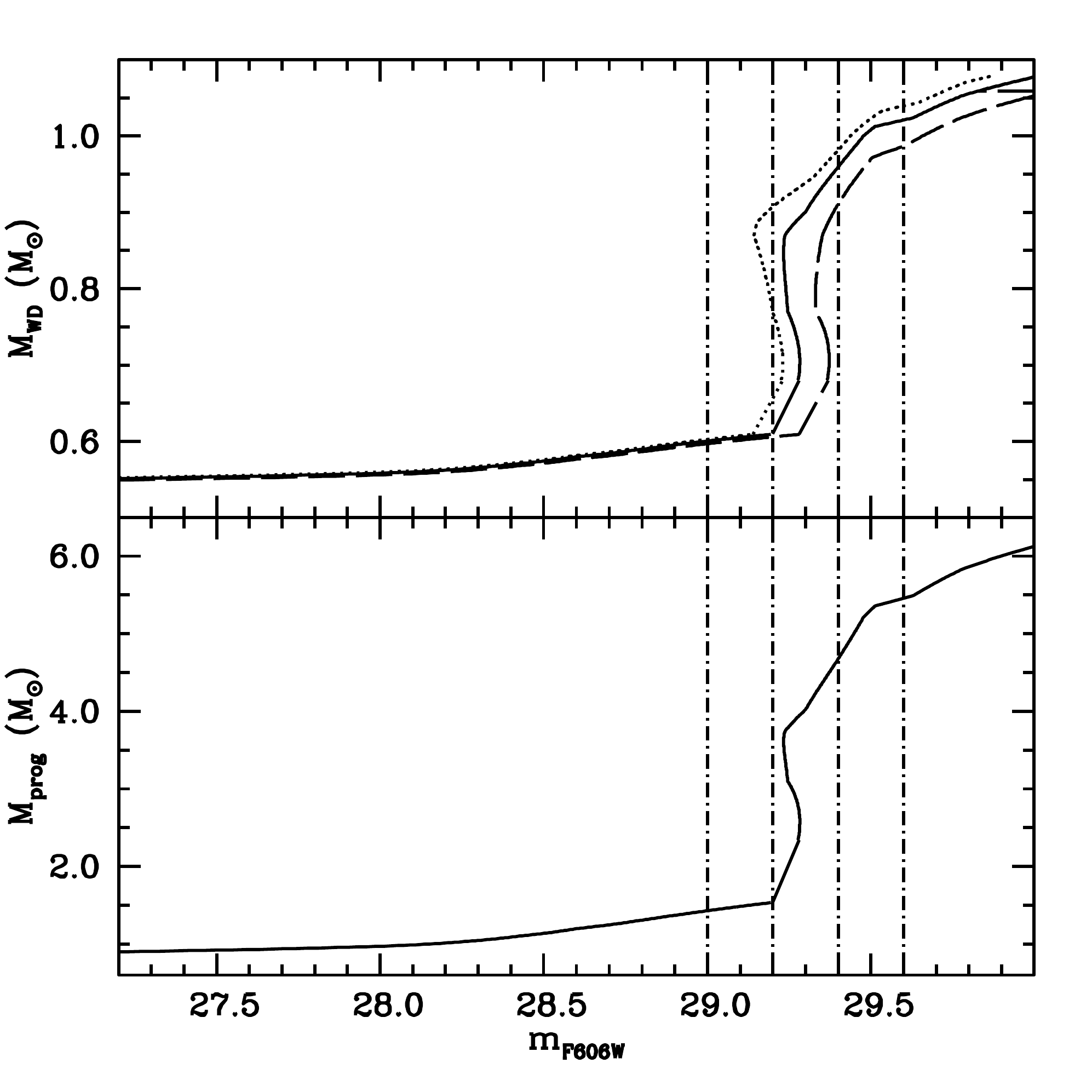}
    \caption{Distribution of the WD masses (in solar mass units -- top
      panel) as a function of the $m_{\rm F606W}$ for the LFs
      displayed in Fig.~\ref{fig:lfage} (keeping the same line-type
      code for the different ages).  The bottom panel displays the
      corresponding distribution of the initial progenitor masses (in
      solar mass units) along the 13~Gyr LF.  The dash-dotted vertical
      lines in the two panels mark the boundaries of the three
      magnitude bins that enclose the region of the peak and
      subsequent cut-off of the empirical LF.}
    \label{fig:massiso}
\end{figure}
%%%

This magnitude range of the theoretical LF is populated by objects
with mass --and progenitor mass-- increasing only very slowly with
magnitude, the average WD mass being around 0.56$M_{\odot}$, as shown
in Fig.~\ref{fig:massiso}.  The number of stars per magnitude bin
will, therefore, be mainly determined by the local cooling speed, with
a smaller contribution from the choice of the progenitor mass
function.  Starting from $m_{\rm F606W} \sim$28.0 the cooling speed
slows down because of the onset of crystallization, hence the local
increase of the number counts.

Beyond $m_{\rm F606W}$=29.0 both theoretical and observed LFs display
a steep increase of the number counts that peak at $m_{\rm
  F606W}$=29.3, followed by a sharp drop at fainter magnitudes. In
this magnitude range the LF is populated by all other more massive
WDs, originated from progenitors with initial mass above
$\sim$1.5$M_{\odot}$, as shown in Fig.~\ref{fig:massiso}.
These objects have reached fainter magnitudes because of their longer
cooling times (shorter progenitor lifetimes).  Their piling up in a
relatively narrow magnitude range explains the appearance of the peak
and cut-off in the LF at the bottom of the CS. As a consequence of
this pile-up of WDs originated by progenitors with a large range of
initial masses, a variation of the exponent of their mass function can
have a major impact on the predicted WD number counts in this
magnitude range.

As an aside, we note that Fig.~\ref{fig:massiso} shows a non-monotonic relation between WD (and progenitor) mass and magnitude at the faint end of the theoretical LF. This is a consequence 
of the fact that at these luminosities the model cooling times do not increase monotonically with increasing WD mass.
This is due to the interplay of the onset (in terms of luminosity) of crystallization that depends on the WD mass (higher masses start crystallizing earlier), and the associated time delays, which are in turn dependent on the 
WD mass (because of the different CO profiles) and  luminosity (energy released at 
higher luminosities induces shorter time delays).

It is very important to notice also that the shape of the WD LF
NGC\,6752 is almost identical to the shape of the LF of the WD CS of
M\,4 \citep[see Fig.~10 in][]{M4}, despite the different metallicity
and HB morphology (redder) of this more metal rich cluster. In M\,4
the WD LF also displays a very mild increase of the number counts
toward fainter magnitudes, followed by a sharper increase and a
roughly flat portion, before the pile up at the bottom end of the CS.
These general trends can also be seen in the WD LFs of the metal poor
cluster NGC\,6397 (with a blue HB morphology) and the red HB, metal
rich cluster 47\,Tuc \citep[see Fig.~2 in][]{hTuc} even though in this
latter cluster the magnitude range spanned by the roughly flat portion
of the LF is narrower, likely because of its younger age.

%%%
\begin{figure}
    \centering
    \includegraphics[width=\columnwidth]{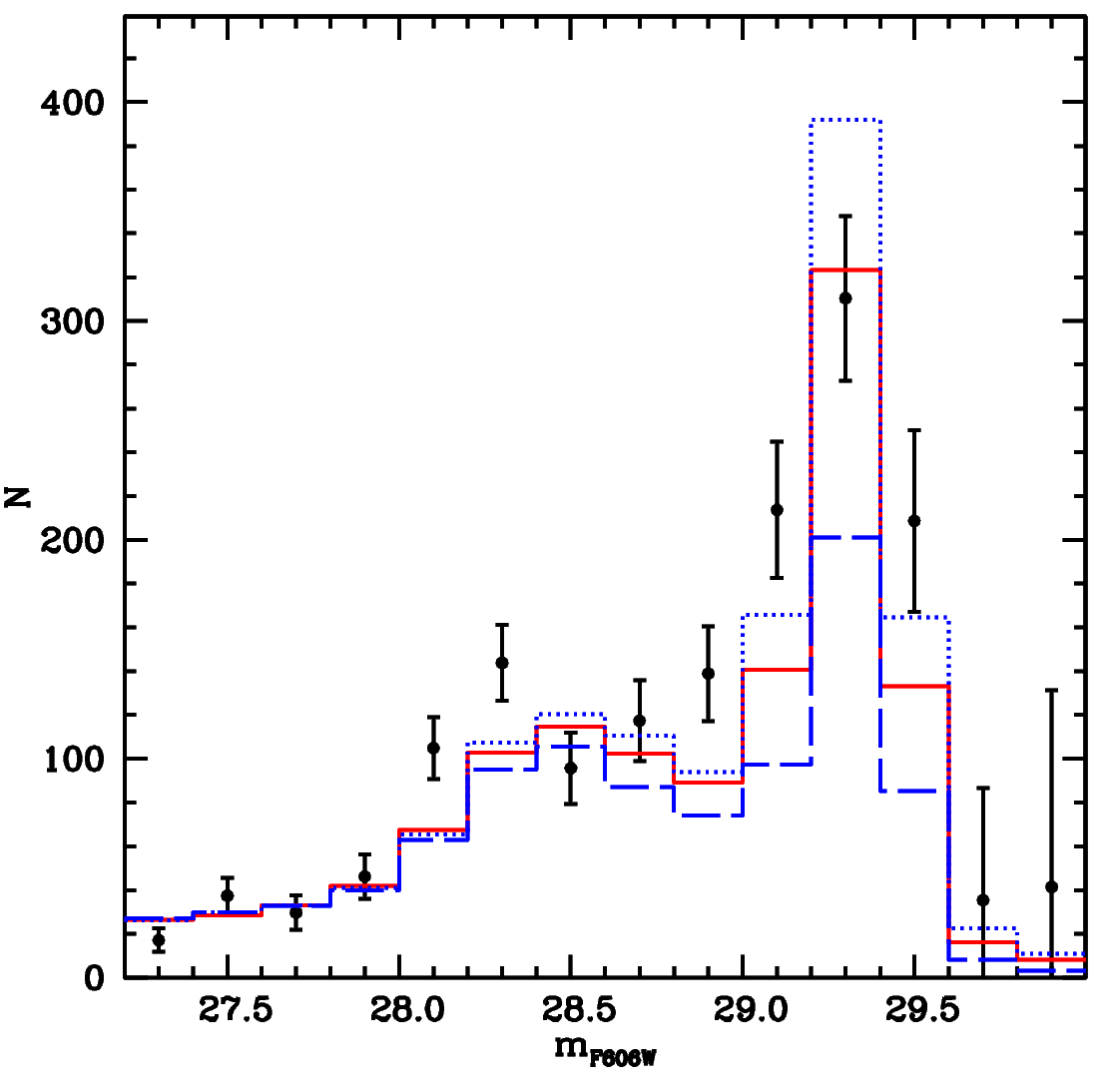}
    \caption{As Fig.~\ref{fig:lfage}, but showing three theoretical
      LFs for an age of 13\,Gyr, calculated with exponents of the
      progenitor mass function equal to $x=-$2.1 (dotted line), $-$2.3
      (solid line, the same LF of Fig.~\ref{fig:lfage}) and $-$2.8
      (dashed line) respectively.}
    \label{fig:lfmf}
\end{figure}
%%% 

One important aspect to consider in the interpretation of the WD LF is
the role of the progenitor mass function and the possible complexities
associated with the effects of the cluster dynamical evolution on this
mass function. In particular, as a result of the interplay between
internal dynamical processes and the external Galactic tidal field,
globular clusters gradually lose stars that escape beyond the
cluster's tidal radius. This mass loss preferentially affects low-mass
stars and may significantly alter the cluster's global mass function
\citep[see, e.g.,][]{vh,bm}. This process may therefore plays a role
in determining the present-day mass distribution of WDs by altering
the mass function of their progenitors before the WD formation and,
more directly, by causing the preferential loss of low-mass WDs after
their formation.  In addition to these effects, it is important to
consider the gradual segregation of massive stars towards the
cluster's central regions and the outward migration of low-mass stars
driven by the effects of two-body relaxation. Since observations do
not typically cover the entire radial extension of a cluster, a
detailed model of the WD mass distribution would also require proper
consideration of the effects of segregation and local variations of
their mass distribution.  Modeling the combined effects of these
processes is non-trivial and would require an extensive suite of
simulations specifically aimed at reconstructing the dynamical history
of the cluster, and at exploring the implications for the mass
distribution of its WD population. This is beyond the scope of this
paper, but we emphasize that part of the discrepancies between the
observed and the theoretical WD LFs shown below might be ascribed to
the dynamical effects we have discussed.

%%%
\subsection{The impacts of the cluster's stellar mass function and initial-final mass relation}
In order to provide a general illustration of the dependence of the WD
LF on the stellar MF, in Fig.~\ref{fig:lfmf} we show the effect of
changing the exponent $x$ of the progenitor's mass function in the
calculation of the 13~Gyr LF.
The magnitudes of peak and cut-off are not affected, but the number
counts are. An exponent $x$=$-$2.15 matches better the observations
compared to the $x$=$-$2.3 case. The total number of stars in the
three bins centred at $m_{\rm F606W}$=29.1, 29.3 (the magnitude of the
peak of the observed LF) and 29.5 of the empirical LF is 733$\pm$64,
where the 1$\sigma$ error has been calculated by propagating the
1$\sigma$ errors on the number counts of the individual bins; the
13~Gyr theoretical LF calculated with $x$=$-$2.1 has 723 objects in
this same magnitude range -- consistent with the observations within
less than 1$\sigma$-- whilst only 600 objects populate the LF
calculated with $x$=$-$2.3.  In the range between $m_{\rm F606W}$=28.1
and 28.9 the empirical LF has 601$\pm$38 objects, whereas the
theoretical counterpart calculated with $x$=$-$2.1 has 497 objects
(477 objects in the LF calculated with $x$=$-$2.3), an underestimate
of more than 2.5$\sigma$.  This latter discrepancy might hint at some
significant underestimate of the cooling timescales of models in this
magnitude range; in fact, a qualitatively similar discrepancy along
the flat part of the LF before the pile up at the end of the CS can be
seen in the comparison of theoretical LFs with observations in M\,4
\citep[see Fig.~13 in][]{M4}.

%%%
\begin{figure}
    \centering
    \includegraphics[width=\columnwidth]{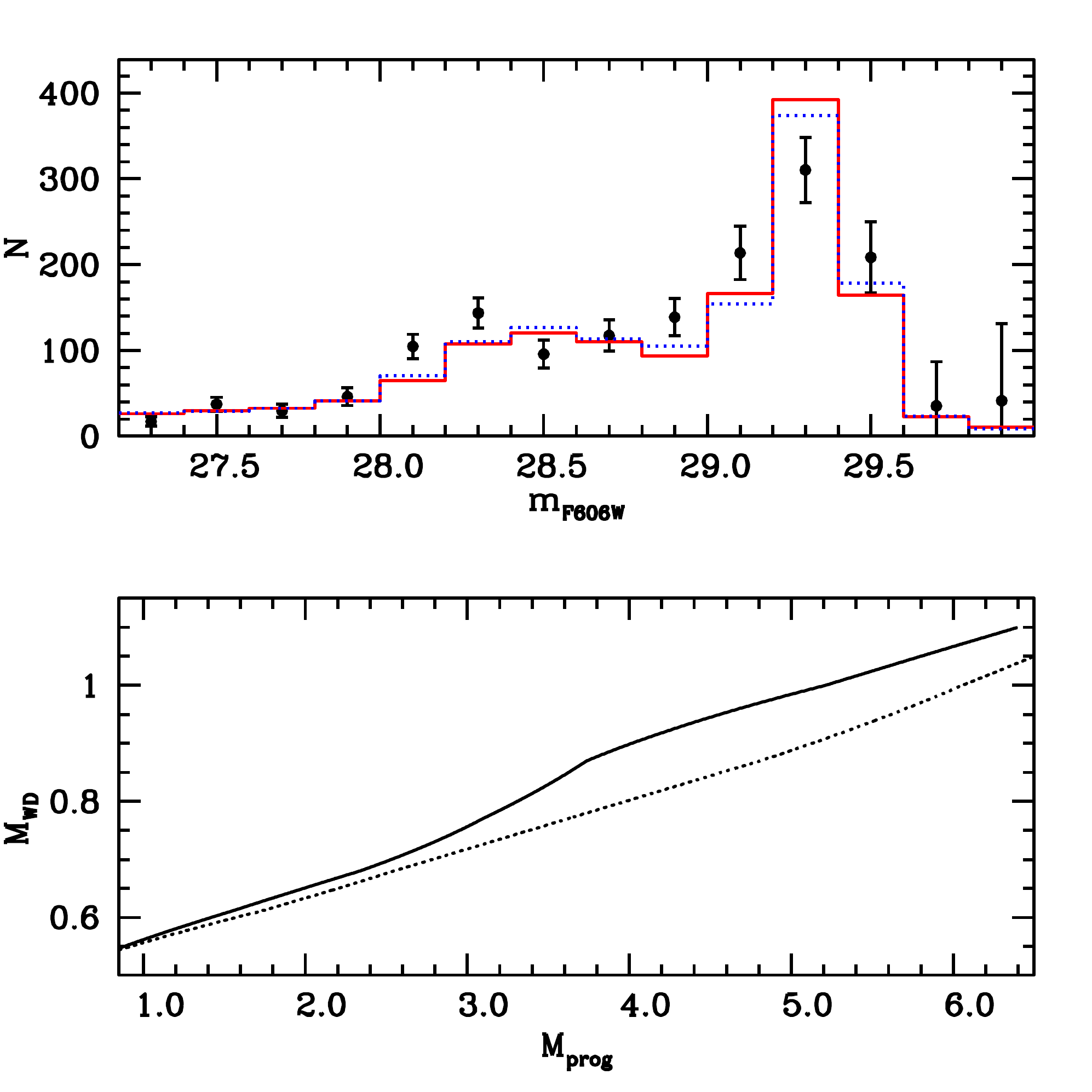}
    \caption{The top panel is as Fig~\ref{fig:lfmf} but showing the
      13~Gyr theoretical LF calculated with $x=-$2.1 (solid line) plus
      a coeval one calculated using the \citet{s09} and $x$=$-$1.95
      (dotted line). The bottom panel compares the IFMRs used in the
      calculation of the LFs in the top panel.}
    \label{fig:lfmimf}
\end{figure}
%%%

In addition to the stellar mass function and its evolution driven by
cluster dynamics, the IFMR may also play a key role in determining the
WD LF. In Fig.~\ref{fig:lfmimf} we display the results of another
numerical experiment. It shows the 13~Gyr LF of Fig.\,~\ref{fig:lfmf}
calculated with $x$=$-$2.1, plus a coeval LF computed by employing
this time the linear IFMR by \citet{s09} and a progenitor mass
function with exponent $x$=$-$1.95.  The two theoretical LFs are
almost identical, despite the different values of $x$ and choice of
IFMR. The number of objects around the peak of the observed LF is 707
for the LF calculated with the IFMR by \citet{s09}, whilst in the
range between $m_{\rm F606W}$=28.1 and 28.9 this LF contains 527
objects, a number that is within 2$\sigma$ of the observations.

To summarize, the quality of the agreement between observed and
theoretical star counts along the LF depends somehow on the choice of
both the progenitor mass function and of the IFMR, but the magnitudes
of the LF peak and cutoff are much more solid prediction of
theory. This confirms that the consistency between MS TO and CS ages
is robust.\\

%%%
\subsection{The impact of the electron conduction opacities}
The empirical WD LF allows us to test also WD models calculated with
the recent electron conduction opacities by \citet{blouin}. These
authors have published improved calculations for H and He compositions
in the regime of moderate degeneracy, which they have combined with
the \citet{opacond} calculations to include the regime of strong
degeneracy and cover the full parameter space necessary to stellar
modelling. The opacities at the transition between moderate and strong
degeneracy are still uncertain \citep[see, e.g.,][]{blouin, opaconf},
but they are crucial for modelling WD envelopes and predicting the
correct cooling times. As investigated by \citet{opaconf}, different
ways to model this transition region give a spectrum of values of the
opacity that vary by up to a factor of $\sim$2.5 in the regime
relevant to WD envelopes.

\begin{figure}
    \centering
    \includegraphics[width=\columnwidth]{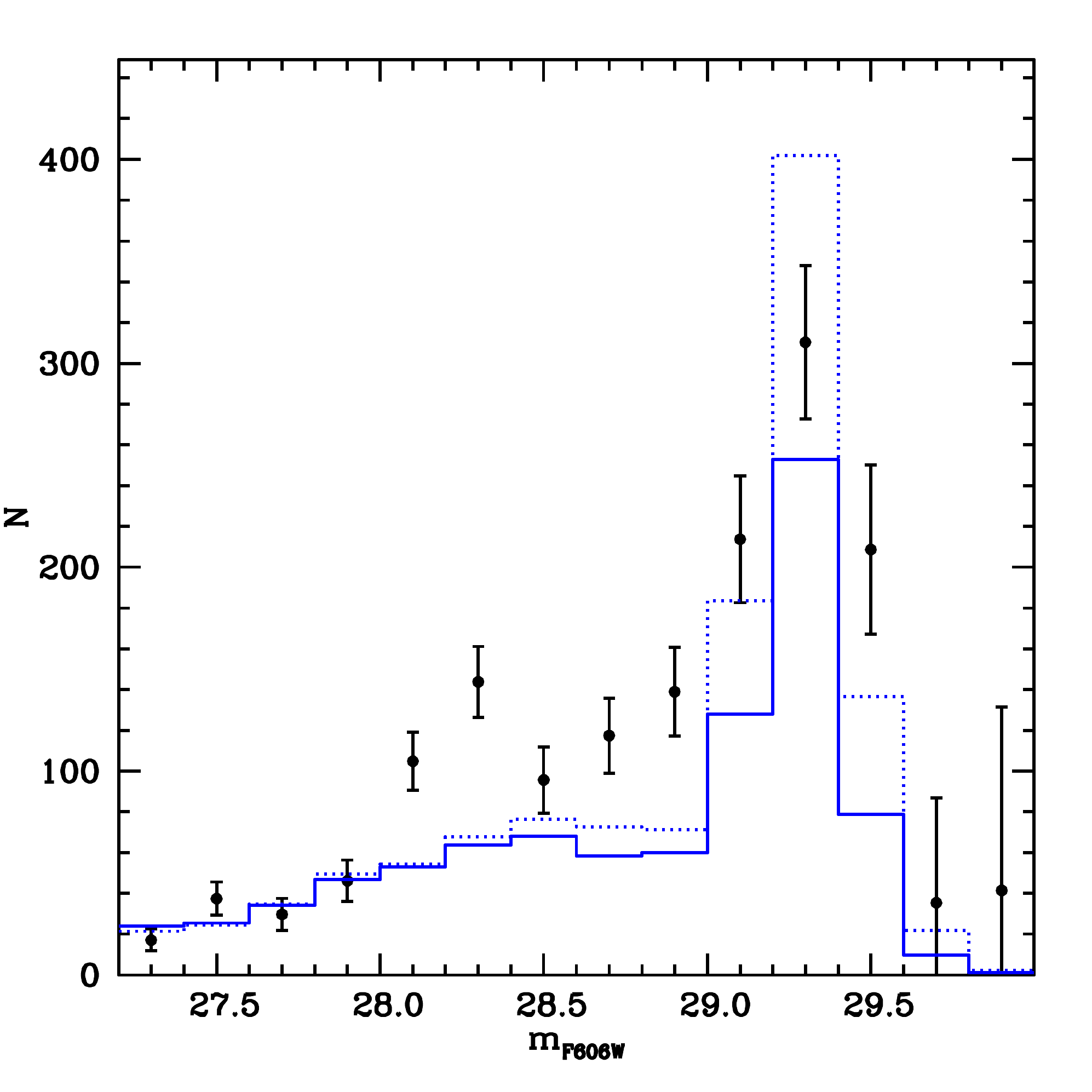}
    \caption{As Fig~\ref{fig:lfage} but showing two 10.7~Gyr
      theoretical LFs calculated using the \citet{blouin} electron
      conduction opacities, and exponents of the progenitors mass
      function equal to $-$2.3 (solid line) and $-$1.8 (dotted line)
      respectively.}
    \label{fig:lfopa}
\end{figure}

We have calculated WD isochrones and LFs as discussed before, this
time employing the set of \citet{bastiwd} WD models computed with the
\citet{blouin} conductive opacities. Figure~\ref{fig:lfopa} compares
the observed LF with two of these new theoretical LFs for an age of
10.7~Gyr, computed with two different exponents of the progenitor mass
function ($x=-$1.8 and $x=-$2.3, respectively).  The age required to
match the observed magnitude of the peak of the LF is now in
disagreement with the age from the MS TO.  Also the number counts in
the magnitude range between $m_{\rm F606W}\sim$28 and $\sim$29, are
largely underestimated, even for the LF calculated with $x$=$-$1.8,
that compares well with the total number of objects around the peak of
the empirical LF (721 objects in the theoretical LF). This is all
consistent with the fact that, apart from the early stages of cooling,
models calculated with \citet{blouin} opacities evolve much faster
than models computed with \citet{opacond} opacities
\citep[][]{bastiwd}.
%%%
%%%

%%%
\begin{figure}
    \centering
    \includegraphics[width=\columnwidth]{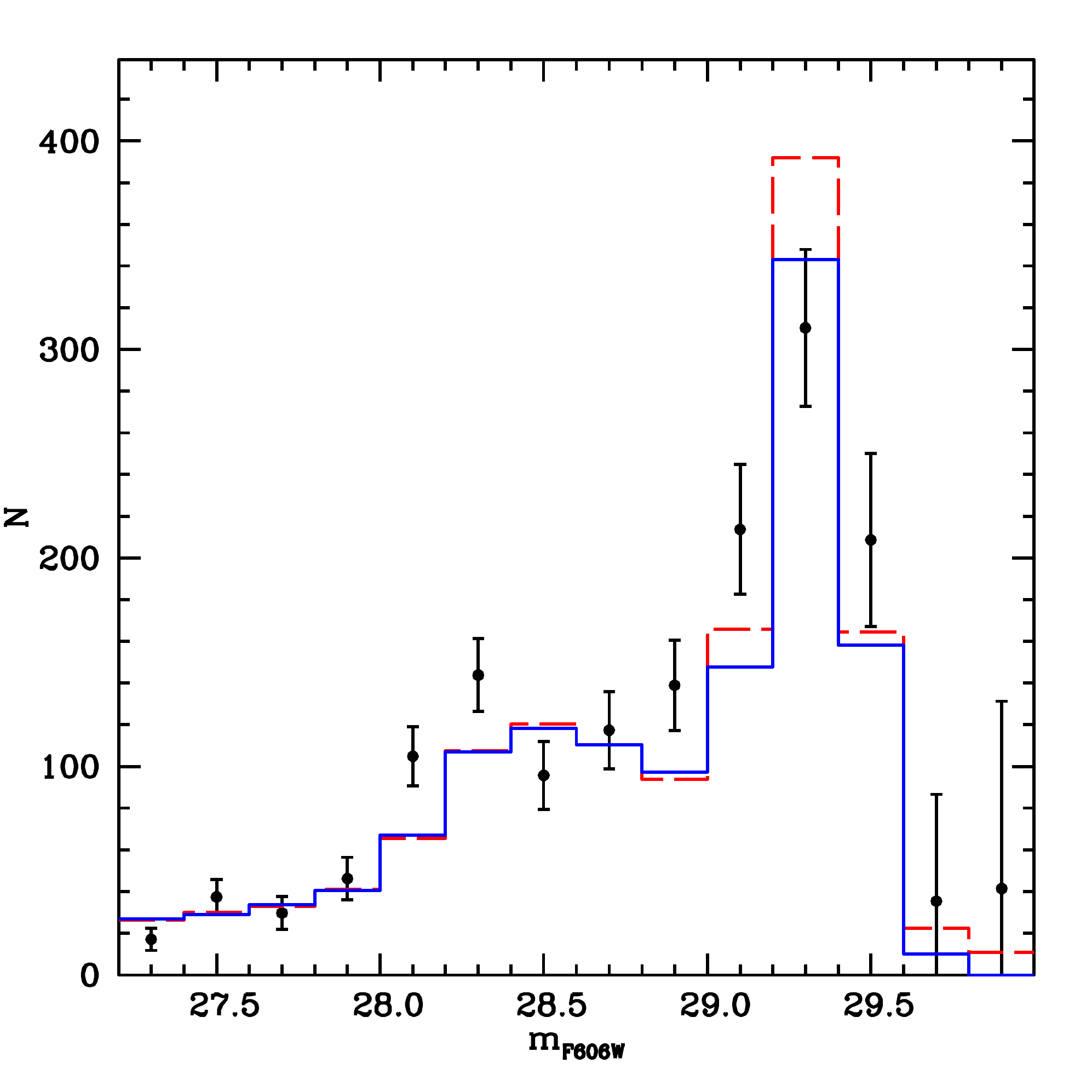}
    \caption{As Fig.~\ref{fig:lfage}, but showing two theoretical LFs
      for an age of 13~Gyr. The dashed line displays a LF calculated
      with helium-normal progenitors and mass function exponent equal
      to $-$2.1 (the same LF displayed as a solid line in
      Fig.~\ref{fig:lfmimf}), whilst the solid line shows a
      calculation with the same mass function but an initial $Y$=0.275
      for the progenitors.}
    \label{fig:LFhe}
\end{figure}
%%%

%%%
\subsection{The impact of the cluster's multiple populations}
We have discussed already how the presence of multiple populations in
the cluster impacts the bright part of the CS, because of the slowly
cooling WDs supported by hydrogen burning in their envelopes, that are
produced by the bluer HB progenitors \citep[see][]{chen22}.  In the
magnitude range we are studying here, these are no longer present, but
the existence of multiple populations can still potentially have an
impact.

In this cluster we have a $\sim$70\% fraction of stars presently
evolving in pre-WD phases with an initial value of $Y$ larger by
typically $\Delta Y$=0.02-0.03 compared to the remaining 30\% of stars
\citep{m17,m18}.  When we assessed the impact of this second
population on the MS TO ages, we calculated WD isochrones and LFs
using progenitor lifetimes from \citet{bastialpha} models with initial
$Y$=0.275. In the assumption that the IFMR is unaffected by this small
variation of the initial He of the progenitors, we have verified with
test calculations that the CO stratification of the WD models is also
insensitive to such a small $\Delta Y$.
Figure~\ref{fig:LFhe} shows how such a small helium enhancement in the
progenitors has a minor impact on the resulting WD LF. For a fixed age
and exponent of the progenitor mass function, after normalization the
second population WD LF is extremely similar to the LF of first
population WDs. Differences are minimal around the peak of the LF,
hence comparisons of the empirical LF with that of a composite
population made of first and second population WDs will give results
essentially equivalent to the case of using models for just first
population WDs.

%%%%%%%%%%%%%%%%%%%%%%%%%%%%%%%%%%%%%%%%%%%%%%%%
%
\subsection{The impact of helium envelope WDs}
\label{sec:He}
\begin{figure}
    \centering
    \includegraphics[width=\columnwidth]{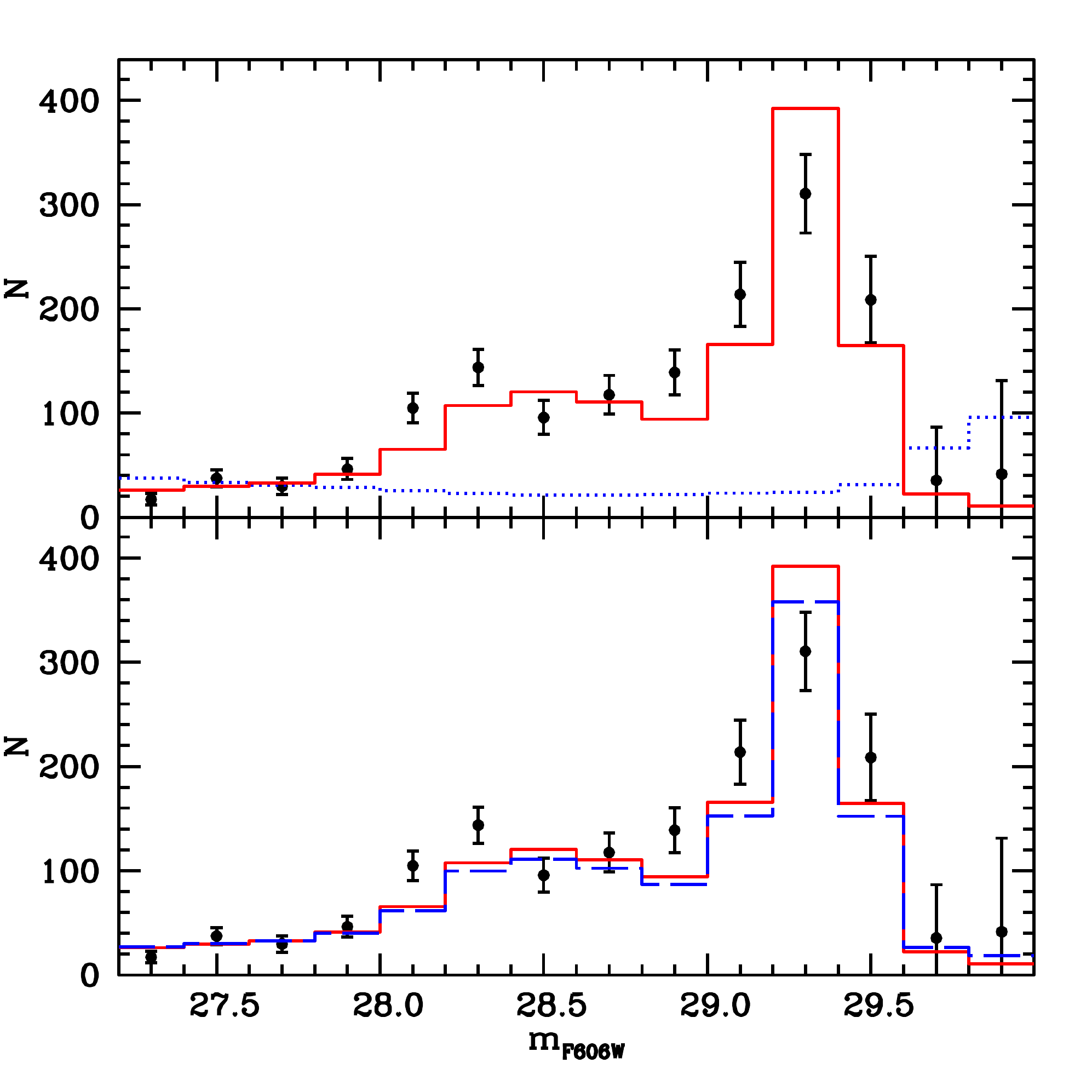}
    \caption{As Fig.~\ref{fig:lfage}, but showing in the top panel
      with a solid line the same 13~Gyr LF plotted as a solid line in
      Fig.~\ref{fig:lfmimf}, populated by WDs with hydrogen envelopes
      and progenitor mass function with $x$=$-$2.1. The dotted line
      denotes a 13~Gyr LF calculated with the same IFMR and progenitor
      mass function but using WD models with helium envelopes (see
      text for details). The lower panel shows the same LF for
      hydrogen envelope WDs (solid line) and the LF for a composite
      population made of both hydrogen and helium envelope WDs (dashed
      line). The number ratio of hydrogen to helium envelope objects
      at formation is set to 90:10 at birth (see text for details).}
    \label{fig:DB}
\end{figure}

%
%%%%%%%%%%%%%%%%%%%%%%%%%%%%%%%%%%%%%%%%%%%%%%%%
%
So far we have studied the WD LF using only models with hydrogen
envelopes.  The top panel of Fig.~\ref{fig:DB} compares with
observations the 13 Gyr LF populated by hydrogen envelope models and
progenitor mass function exponent $x=-$2.1 of Fig.~\ref{fig:lfmimf},
and a LF populated by helium envelope models. This latter LF had been
computed as described before, using \citet{bastiwd} models with pure
helium envelopes.
\footnote{We notice that \citet{nda} have shown how for WDs in the
  \textit{Gaia} catalogue with helium dominated atmospheres, the
  presence of a small percentage of hydrogen is required to determine
  more accurate stellar parameters.}

The shape of the LF populated by helium envelope models is very
different from the observed one. It stays essentially flat from the
magnitude range chosen for the normalization down to $m_{\rm
  F606W}\sim$29.5, completely at odds with observations. Due to the
generally faster cooling times of helium envelope models, in this
magnitude range this LF is populated by objects with practically
constant mass, around 0.55$M_{\odot}$, with progenitor initial masses
in a relatively narrow mass range (the more massive objects appear at
much fainter magnitudes). This also means that the number counts are
insensitive to the choice of the exponent of the progenitor mass
function.

The comparison shows clearly that the population of helium envelope
objects must be a relatively small fraction of the total number of WDs
observed, otherwise the shape of the cluster LF would be very
different from what is observed\footnote{This is strictly true in the
  hypothesis that hydrogen envelope WDs do not transform to helium
  envelope objects during their cooling evolution, because of
  convective mixing with the underlying more massive helium layers
  \citep[see, e.g., the discussion in][]{davis}.}.

On the observational side, \citet{moehler} have studied
spectroscopically 5 objects along the cluster CS with $T_{\rm eff}$
between $\sim$20,000~K and $\sim$13,000~K (corresponding to $m_{\rm
  F606W}$ between $\sim$24 and 24.7 according to our isochrones) and
found they all have hydrogen atmospheres.  \citet{rb} found that in
their sample of 21 bright cluster WDs with $T_{\rm eff}$ between
$\sim$20,000~K and $\sim$10,000\,K, the positions of two objects in the
CMD is consistent with helium atmosphere WDs ($\sim$10\%).

We also note that the faint end of the observed CS in
Fig.~\ref{fig:isofit} has a width larger than what predicted from the
artificial star analysis. This could be hinting at the presence of
helium envelope WDs, that towards the faint end of the CS are expected
to become increasingly redder than their hydrogen envelope
counterpart, as shown in Fig.~\ref{fig:isofit}.

Assuming that the \citet{rb} result represents the true ratio between
hydrogen and helium envelope WDs at birth (9:1), these WDs populate a
magnitude range within the limits of =the \citet{chen22} study, where
the number ratio \lq{slow\rq}/\lq{canonical\rq} WDs is equal to $\sim
70/30$. Our adopted WD models show that in the magnitude range of
\citet{chen22} analysis the 0.54$M_{\odot}$ (a representative mass
populating this temperature range) models with hydrogen and helium
envelopes have the same cooling times, implying that \citet{chen22}
results obtained considering only hydrogen envelope models are
unaffected if we include this small fraction ($\sim$10\% of the total)
of helium envelope objects in the 30\% of \lq{fast\rq} WDs without
hydrogen burning.

For our purposes we have calculated a LF made of a composite
population with 90\% hydrogen envelope objects, and 10\% helium
envelope WDs at birth, both for an exponent of the progenitor mass
function $x$=$-$2.1, assuming there is no transformation from
hydrogen- to helium envelopes during the cooling.  The lower panel of
Fig.~\ref{fig:DB} replaces the LF of helium envelope models with this
composite population in the magnitude range of interest to our
analysis.  Due to the generally faster cooling times of the helium
envelope models, they make only a few percent of the total number of
objects within the magnitude bins centred between $m_{\rm F606W}
=$\,27.3 and 29.5, and the composite LF is almost identical to the one
calculated with just hydrogen envelope WDs.

%%%%%%%%%%%%%%%%%%%%%%%%%%%%%%%%%%%%%%%%%%%%%%%%
%%%%%%%%%%%%%%
\section{Summary}
\label{conclusions}
This is our conclusive study on the WD\,CS of NGC\,6752, which was the
main goal of the multi-cycle \textit{HST} large program
GO-15096/15491.
With respect to Paper\,III, which was based only on the first half of
the data, we now double the total exposure time, and extended our work
taking advantage of all the data collected to better define the
cluster CS and its LF.  This has allowed us to perform robust and
conclusive comparisons with theory; more specifically, the WD models
by \citet{bastiwd}.

We found that the shape of cluster WD LF is very similar to its
counterpart in the metal richer, redder horizontal branch GC M\,4
\citep{M4}, and that theoretical LFs for hydrogen envelope WD models
calculated with the \citet{opacond} electron conduction opacities
follow closely the observed trend of star counts as a function of
magnitude.  The observed LF peaks at $m_{\rm F606W} \simeq 29.3 \pm
0.1$, consistent within uncertainties with what has been previously
reported, and the magnitude of this peak is matched by WD isochrones
with ages between 12.7 and 13.5 Gyr, consistent with the cluster age
derived from the MS TO and subgiant branch.  We confirm that the
predicted magnitude of the LF peak and cutoff is unaffected by
realistic variations of the adopted IFMR and the progenitor mass
function.  We also find that the impact of the cluster multiple
populations on the WD LF for $m_{\rm F606W}$ larger than $\sim$27.3 is
negligible.

Our analysis also reveals a possible hint of an underestimate of the
cooling timescales of models in the magnitude range between $m_{\rm
  F606W}$=28.1 and 28.9. However, different choices of the IFMR can
reduce the discrepancy between theoretical and empirical star counts
in this magnitude range to below 2$\sigma$.  A hint of a similar
discrepancy can be found in the analysis of the WD LF in M\,4 by
\citet{M4}.

\citet{rb} photometry of a small sample of bright cluster WDs suggests
the presence of a small fraction ($\sim$10\%) of WDs with helium
envelopes.  We find that this fraction of helium envelope objects has
a negligible impact on the shape of the LF, and is potentially
responsible for the colour width of the faint end of the observed CS.

Finally, we find that hydrogen envelope models calculated with the
\citet{blouin} electron conduction opacities provide WD ages in
disagreement (they are too low) with the TO age.

%%%%%%%%%%%%%%
\section*{Acknowledgements}
This study is based on observations with the NASA/ESA 
\textit{Hubble Space Telescope}, obtained at the Space Telescope Science Institute,
which is operated by AURA, Inc., under NASA contract NAS 5-26555.
The authors LRB, MSc, MG and DN acknowledge support by MIUR under PRIN
program \#2017Z2HSMF and by PRIN-INAF\,2019 under program \#10-Bedin.
JA, ABu, RG, DA, ABe and RMR acknowledge support from HST-GO-15096 and
HST-GO-15491.
MSa acknowledges support from The Science and Technology Facilities
Council Consolidated Grant ST/V00087X/1.

%%%%%%%%%%%%%%%%%%%%%%%%%%%%%%%%%%%%%%%%%%%%%%%%%%
\section*{Data Availability}

The data underlying this article were accessed from the Mikulski
Archive for Space Telescopes (MAST), available at
\url{https://archive.stsci.edu/hst/search.php}.
All data come from \textit{HST} programs GO-15096 and GO-14662 (on
both P.I. Bedin).
The full list of observations are reported at these two urls following
the archival pages:
%%% 
\url{www.stsci.edu/cgi-bin/get-proposal-info?id=15491}
\url{www.stsci.edu/cgi-bin/get-proposal-info?id=15096}
following the link \texttt{HST Archive Information}.  
%%%%%
The stellar models employed in this article are available at 
\url{http://basti-iac.oa-abruzzo.inaf.it/}

As supplementary online material for the present work, 
we also release the observed LF, the corrected LF, the 
completeness, the photometric errors, and the observed 
CMD. 
%

\bibliographystyle{mnras}
\bibliography{bibliography} % if your bibtex file is called example.bib
%

%\begin{thebibliography}{99}
%\end{thebibliography}

%%%%%%%%%%%%%%%%%%%%%%%%%%%%%%%%%%%%%%%%%%%%%%%%%%

%%%%%%%%%%%%%%%%% APPENDICES %%%%%%%%%%%%%%%%%%%%%

% \appendix

% \section{Some extra material}

% If you want to present additional material which would interrupt the flow of the main paper,
% it can be placed in an Appendix which appears after the list of references.

%%%%%%%%%%%%%%%%%%%%%%%%%%%%%%%%%%%%%%%%%%%%%%%%%%

% Don't change these lines
\bsp	% typesetting comment
\label{lastpage}
\end{document}